\documentclass[conference]{IEEEtran}
\IEEEoverridecommandlockouts
\usepackage{amsmath,amsfonts}
\usepackage{algorithmic}
\usepackage{algorithm}
\usepackage{amsmath}
\usepackage{array}
\usepackage{textcomp}
\usepackage{stfloats}
\usepackage{url}
\usepackage{verbatim}
\usepackage{tikz}
\usepackage{cite}
\usepackage{duckuments}
\usepackage{fancyhdr}
\usepackage{lipsum} 
\usepackage{float} 
\usepackage{xcolor}
\usepackage{graphicx}
\usepackage{pifont}
\usepackage{subfigure}
\usepackage{xspace}
\usepackage{authblk}
\usepackage{placeins}

\ifodd 0
\definecolor{darkgreen}{RGB}{0,200,0}
\definecolor{yellow}{RGB}{255,150,0}
\newcommand{\rev}[1]{{\color{black}#1}} 
\newcommand{\newrev}[1]{{\color{red}#1}} 
\newcommand{\needrev}[1]{{\color{darkgreen}#1}} 
\newcommand{\zrev}[1]{{\color{yellow}#1}} 
\else
\newcommand{\rev}[1]{#1}
\newcommand{\newrev}[1]{#1} 
\newcommand{\needrev}[1]{#1} 
\newcommand{\zrev}[1]{#1}
\fi

\newcommand{\name}{DeepRM\xspace}

\begin{document}

\title{Constructing 4D Radio Map in LEO Satellite Networks with Limited Samples}



\author{Haoxuan Yuan\textsuperscript{1,2}, Zhe Chen\textsuperscript{1,2}, Zheng Lin\textsuperscript{1,2}, Jinbo Peng\textsuperscript{1,2}, Yuhang Zhong\textsuperscript{1,2}, Xuanjie Hu\textsuperscript{1,2}, \\Songyan Xue\textsuperscript{3}, Wei Li\textsuperscript{4}, Yue Gao\textsuperscript{1,2}\\Institute of Space Internet, Fudan University, China\textsuperscript{1}\\School of Computer Science, Fudan University, China\textsuperscript{2}\\Huawei Technologies Co Ltd, China\textsuperscript{3}\\The State Radio Monitoring Center, China\textsuperscript{4}}






\maketitle

\begin{abstract}
\rev{
Recently, Low Earth Orbit (LEO) satellite networks (i.e., non-terrestrial network (NTN)), such as Starlink, have been successfully deployed to provide broader coverage than terrestrial networks (TN). Due to limited spectrum resources, TN and NTN may soon share the same spectrum. Therefore, fine-grained spectrum monitoring is crucial for spectrum sharing and interference avoidance. To this end, constructing a 4D radio map~(RM) \newrev{including} three spatial dimensions and signal spectra is \newrev{important}.} However, this requires \newrev{the large deployment of sensors, and high-speed analog-to-digital converters for extensive spatial signal collection and wide power spectrum acquisition, respectively.} To address these challenges, we propose a deep unsupervised learning framework without ground truths labeling requirement, \name, \newrev{comprised of} neural compressive sensing (CS) and tensor decomposition (TD) algorithms. Firstly, \rev{we map the CS process into the optimization of a neural networks-associated loss function, \newrev{and design a sparsity-performance balance training algorithm to reconstruct a wide power spectrum under limited sub-Nquist samples.} 
Secondly, \newrev{according to the output of neural CS algorithm}, we also utilize neural networks to perform TD, and construct the 3D RM for each frequency, even under very sparse sensor deployment.}
\newrev{Extensive evaluations} show that \name achieves lower error than its corresponding state-of-the-art baselines, especially with limited samples.
\end{abstract}

\begin{IEEEkeywords}
LEO satellites, radio map, compressive sensing, tensor decomposition, neural network.
\end{IEEEkeywords}

\section{Introduction}\label{sec:introduction}

In recent years, 
\newrev{as a type of non-terrestrial networks~(NTNs), the low Earth orbit~(LEO) mega-constellations such as Starlink and OneWeb, are deployed successfully
~\cite{zhao2024leo,yuan2023graph,michel2022first,lin2024fedsn,yuan2024satsense}. Compared with terrestrial networks~(TNs), the LEO constellations offer much wider coverage~\cite{xiao2022leo,ma2023network}. 
\zrev{The ever-increasing number of LEO constellations and limited spectrum resources~\cite{liu2018space,cui2022space,yue2023low,lin2021spatial} pose a significant barrier to the democratization of space-air-ground integrated networks~(SAGIN).}
NTN and TN may serve as primary and secondary users to enable spectrum sharing, respectively, since we control and manage TN much more easily than NTN~\cite{whitepaper}.} Therefore, effective spectrum monitoring techniques for NTN, particularly for LEO satellite networks are crucial. 

\begin{figure}[t]
    \centering
    \includegraphics[width=1\linewidth]{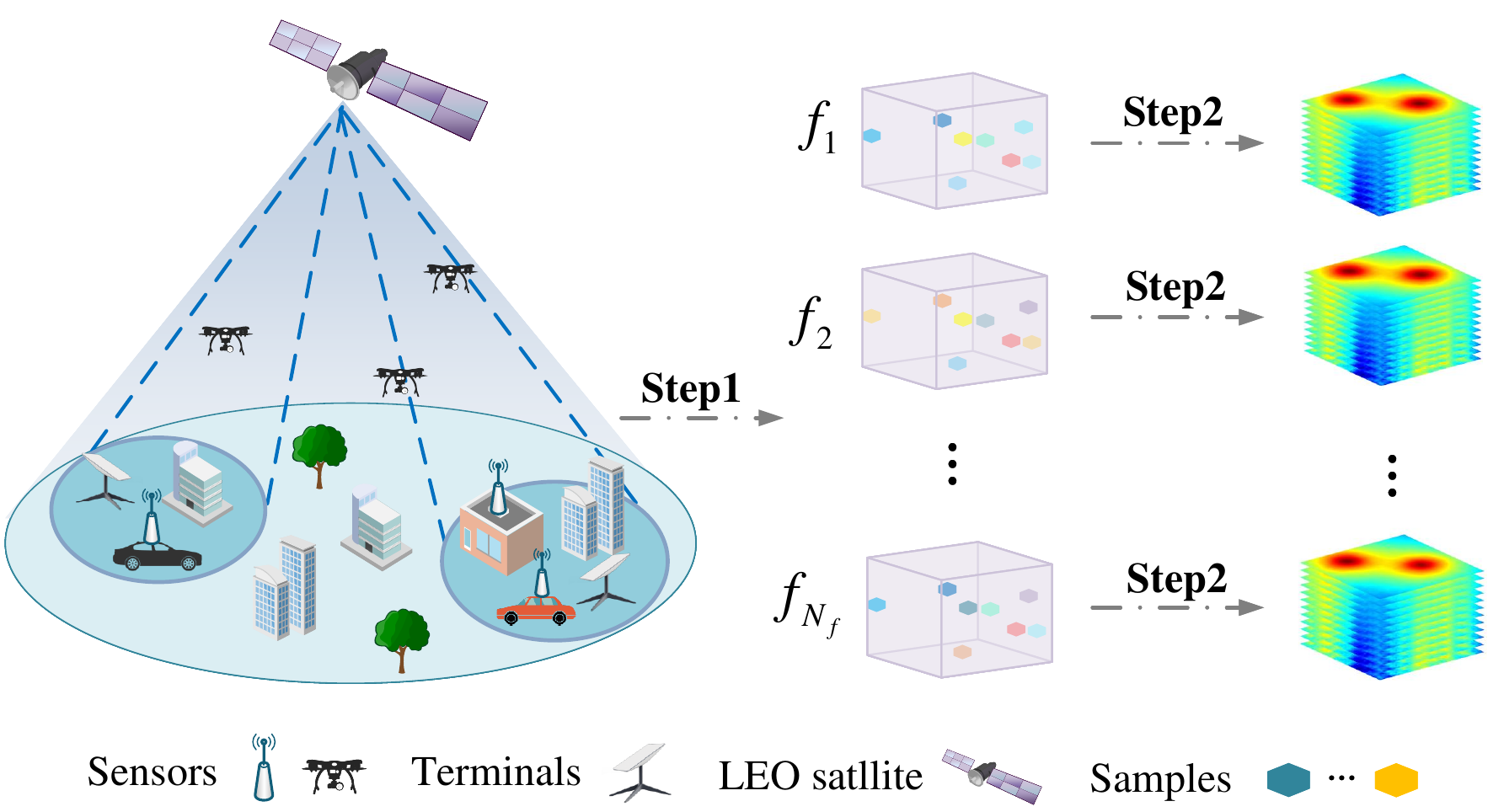}
    \caption{\newrev{The 4D RM constructing process in LEO satellite networks, where $f_i$ denotes the $i$-th frequency of the wideband.}}
    \label{fig:teaser}
\end{figure}

\newrev{Spectrum sensing~(SS) is the conventional spectrum monitoring technique to}
identify the usage of and find idle spectrum\cite{chakraborty2017specsense,uvaydov2021deepsense,luo2023spectrum,peng2024sums}, \newrev{but fails to offer fine-grained spectrum monitoring, due to lack of spatial dimensions. Recent research works upgrade SS to build radio map~(RM) with only a single frequency by collecting received signal strengths~(RSSs) in 2D~\cite{bi2019engineering,zeng2024tutorial} or 3D~\cite{shen20213d,wang2023sparse,liu2024uav,hu20233d} space. Although there are a few works for 2D RM with multiple frequencies~\cite{zhang2020spectrum,shrestha2022deep,teganya2021deep}, missing height information makes it hard to dynamically allocate spectrum resources for unmanned aerial vehicles in SAGIN.}
Consequently, it is essential to accurately \newrev{construct 3D RM with frequency dimension~(a.k.a. \textit{4D RM}) in LEO satellite networks.}
We provide an intuitive representation of the construction process of the 4D RM, as shown in Fig.~\ref{fig:teaser}. In the first step, a large number of aerial and ground sensors are deployed \newrev{to collect RF signal information at different frequencies} in the area of interest, \newrev{and transmit the RF signal information to the central server. In the second step, the center server leverages a construction algorithm to achieve 4D RM. However, in practice, to construct a 4D RM is a non-trivial task, we face two key challenges:}
\begin{itemize}
    \item \newrev{ \textit{Considering frequency dimension, to capture a large bandwidth spectrum requires a high-speed ADC, since the state-of-the-art sweeping algorithms with low sampling rate~\cite{guddeti2019sweepsense} may miss signals in LEO satellite networks.}  
    LEO satellite networks always occupy a large spectrum with multi-band signals. Specifically, \needrev{for downlink, Starlink utilize a single bandwidth with 240\!~MHz bandwidth, and total 8 bands ranging from 10.7\!~GHz to 12.7\!~GHz~\cite{humphreys2023signal}.} Consequently, a new sampling solution for low sampling rates needs to be designed.}
    
    \item \newrev{\textit{Different from the construction of TN 3D RM with a few sensors in small-scale area~(e.g., \needrev{0.01\!~$km^2$})~\cite{shen20213d, wang2023sparse}, we need the large number of sensors to collect signals in LEO satellite networks~\cite{hu20233d,liu2024uav}.} Due to higher satellite-ground distance~(e.g., \needrev{550}\!~$km$), even a narrow beamwidth~(e.g., $1.5^{\circ}$) generated by phased-array can cover very wide area~(e.g., \needrev{160\!~$km^2$~\cite{mcdowell2020low}}). Therefore, the overhead of sensor deployment is really large.
    }
\end{itemize}

To address the above two challenges, we formulate them as a deep neural networks~(NNs) problem, and then, propose a deep unsupervised learning framework, named \name, for 4D RM construction with limited sensors and the sampling rate of ADCs. 

\newrev{For the first challenge, \name does not rely on sweeping algorithms, but employs a multi-coset sub-Nyquist sampling method via leveraging multiple low sampling rate \needrev{ADCs~\cite{song2019real, peng2024sums}} to capture the whole spectrum in LEO satellite networks. The reconstructive algorithms must be applied to sub-Nyquist samples, for original samples recovery. However, the performance of state-of-the-art reconstructive algorithms degrades significantly \rev{when the sampling rate is 1/4 of the Nyquist sampling rate}~(see Sec.~\ref{sec:sub-Nyquist}). We push the limit of reconstruction performance of sub-Nyquist sampling using our neural compressive sensing~(CS) algorithm. We design an NNs solver to represent the variables in the sub-Nyquist sampling CS problem, and a training algorithm to strike a balance between sparsity and performance via loss function. Hereby, the power spectrum of the whole LEO satellite networks is reconstructed successfully with low-speed and low-cost ADCs, and is used to construct 3D RMs.}

\newrev{To combat the second challenge, we propose a neural tensor decomposition algorithm based on highly sparse sensors. We are inspired by the tensor decomposition~(TD), in particular tucker decomposition, but use NNs to represent its factor matrices. Therefore, while we train the NNs to approximate factor matrices as close as possible, they contain richer intrinsic features, due to their nonlinear representation capability. To revert~(i.e., multiply) the trained decomposition NNs, we accurately construct the 3D RM with specific frequency. We summarize our contributions in the following:
}
\begin{itemize}
    \item \newrev{To the best of our knowledge,} this is the first work to construct a 4D RM of LEO satellite networks, offering comprehensive and practical information for satellite spectrum monitoring across spatial and frequency dimensions.
    \item We design a deep unsupervised framework, \name for 4D RM construction, which innovatively leverages the nonlinear fitting capabilities of NNs to address optimization problems in CS and TD, thereby achieving high fidelity reconstruction results with highly limited samples.
    \item Extensive experimental results demonstrate that \name outperforms its corresponding state-of-the-art methods.
\end{itemize}

This paper is organized as follows. Sec.~\ref{sec:motivation} introduces background and motivation for our design. Sec.~\ref{sec:design} presents the system design, and Sec.~\ref{sec:implementation} describes system implementation and experimental setup, followed by performance evaluation in Sec.~\ref{sec:evaluation}. Related works and technical limitations are discussed in Sec.~\ref{sec:related_work}. Finally, conclusion and future remarks are stated in Sec.~\ref{sec:conclusion}.
\section{Motivation and Background}\label{sec:motivation}
In this section, we first elaborate on the components of the 4D RM. Then, we present two challenges encountered in the 4D RM construction process.
\begin{figure}[t]
    \centering
    \includegraphics[width=0.8\linewidth]{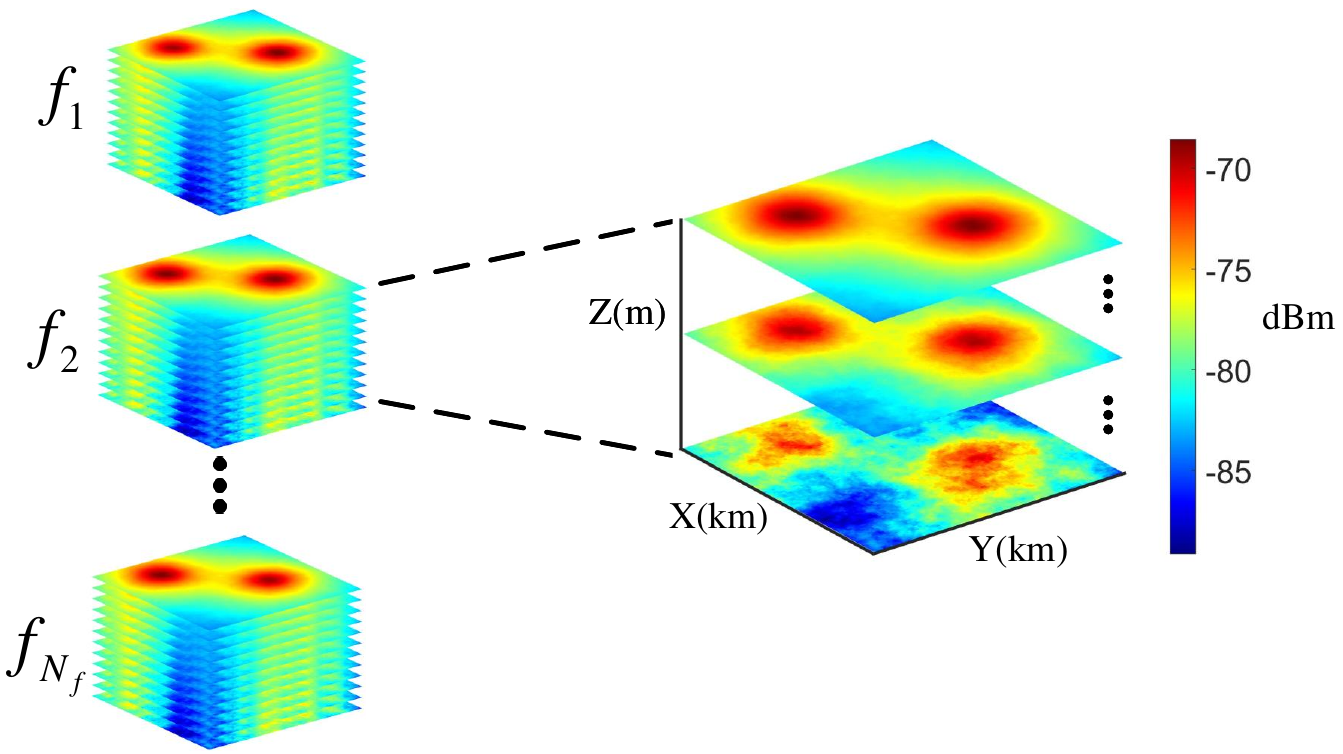}
    \caption{\rev{\newrev{An example} of a 4D RM tensor.}}
    \label{fig:illustration}
\end{figure}

\subsection{4D RM Tensor}\label{sec:4D tensor}

\zrev{Fig.~\ref{fig:illustration} shows the 4D RM tensor, denoted as $\mathbf{\mathcal { T }} \in \mathbb{R}^{N_f \times I \times J \times K}$, where $N_f$ denotes the number of frequency bins, and $I\times J\times K$ represent the size of the 3D space. The 3D RM tensor ${\mathbf{\mathcal { X }}}_i$ at frequency $f_i$ is denoted by ${{\mathbf{\mathcal { X }}}_i} \in \mathbb{R}^{ I \times J \times K}$, with each element in $\mathbf{\mathcal { X }}$ representing the RSS at corresponding 3D spatial positions. From Fig.~\ref{fig:illustration}, It is evident that the upper horizontal slices of the 3D RM correspond to higher vertical altitudes, where the downlink signals from LEO satellites experience minor shadowing effects from obstacles such as buildings and mountains. Consequently, the RSS attenuation is minor, resulting in a relatively smooth distribution of RSS.  However, in the lower horizontal slices, the shadowing effects become increasingly pronounced, causing a more irregular and blurred distribution of the RSS. The details of generating the 4D RM data will be elaborated upon in Sec.~\ref{sec:implementation}. }


\subsection{Sparse Informative-rich Sampling}\label{sec:sub-Nyquist}
\begin{figure}[t]
\centering  
\subfigure[Ground-truth]{
\label{fig:rank}
\includegraphics[width=0.23\textwidth]{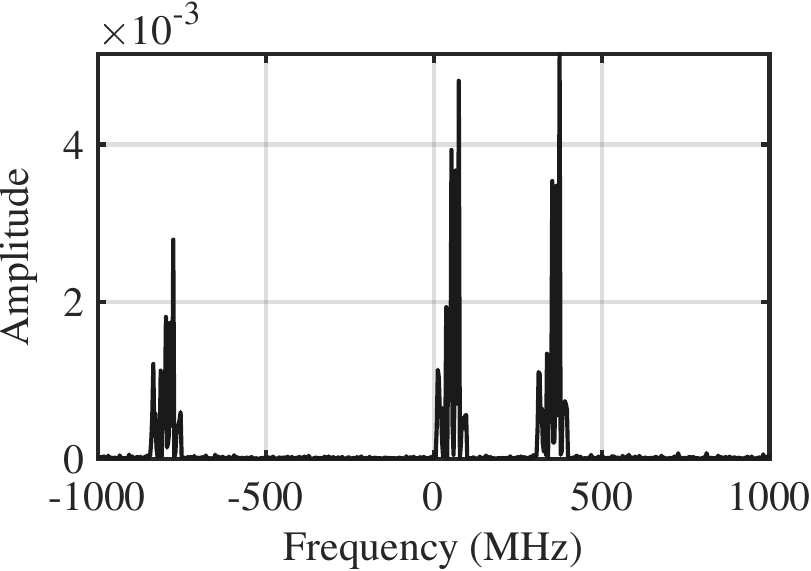}}
\subfigure[FCPSE]{
\label{fig:mse}
\includegraphics[width=0.235\textwidth]{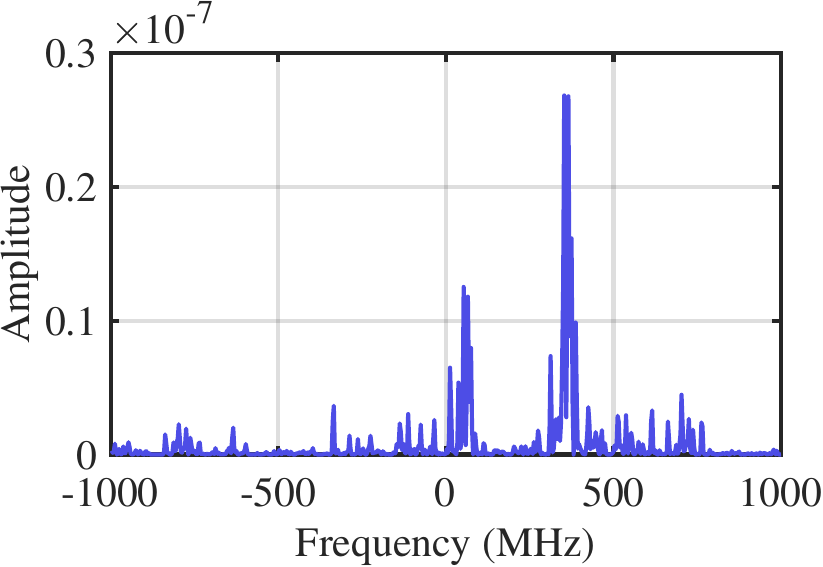}}\\
\subfigure[JB-HTP]{
\label{fig:additional1}
\includegraphics[width=0.23\textwidth]{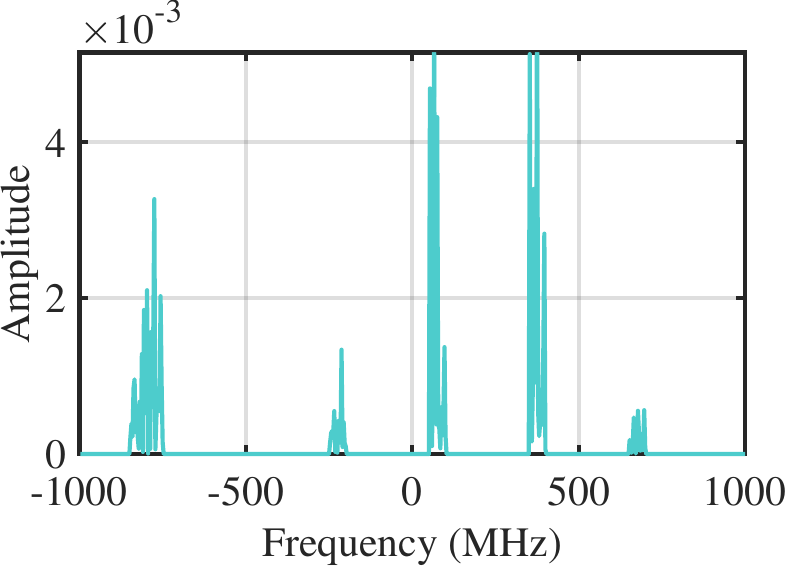}}
\subfigure[SOMP]{
\label{fig:additional2}
\includegraphics[width=0.23\textwidth]{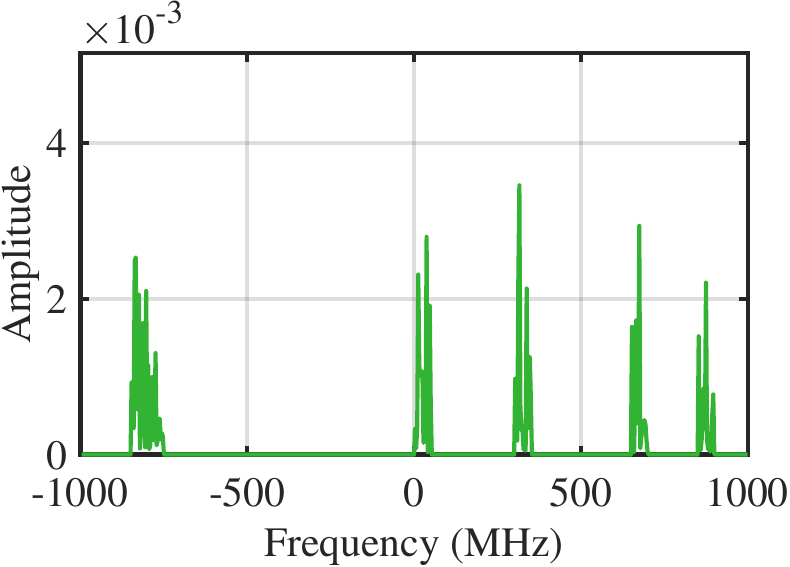}}
\caption{The ground-truth power spectrum (a) and reconstructed power spectrum using FCPSE~\cite{yang2019fast} (a), JB-HTP (b) and SOMP (c)~\cite{song2019real}, where the sampling rate is 1/4 of the Nyquist sampling rate.}
\label{Fig.challenge2}
\end{figure}

\zrev{Recalling Sec.~\ref{sec:introduction}, the construction of 4D RM necessitates sensors equipped with high-speed ADCs to achieve the target sampling rates. CS enables the sampling of signals at sub-Nyquist rates and accurate reconstruction of the power spectrum, indicating that sensors equipped with low sampling rate ADCs can still capture RSS information across a wide range of frequencies under a wideband scenario. Classic power spectrum reconstruction algorithms such as SOMP, JB-HTP~\cite{song2019real} and FCPSE~\cite{yang2019fast} are widely adopted due to their superior performance and relatively low computational complexity. However, these algorithms struggle to effectively reconstruct the power spectrum at extremely low sampling rates. At low sampling rates, the collected data points become sparse and may not capture all the necessary details of the signal’s frequency components. This sparsity leads to missing critical information about the signal, rendering it difficult for algorithms to accurately infer the power spectrum. We conduct a motivating experiment to compare the ground-truth power spectrum and reconstructed power spectrum with FCPSE, JB-HTP, and SOMP under 1/4 of the Nyquist sampling rate. It is seen from Fig.~\ref{Fig.challenge2} that signals reconstructed using SOMP, JB-HTP, and FCPSE differ significantly from the ground-truth signal. This discrepancy highlights a paramount challenge in power spectrum reconstruction at a low sampling rate. Except for the wideband sampling limitation, we will meet the sparse sensors limitation challenge in the following section.}

\subsection{Restricted Sensor Coverage} \label{ssec:ssl}

\zrev{
The extensive coverage range of satellites results in a large 3D RM size in satellite networks. However, monitoring such large areas necessitates the deployment of excessive sensors across a wide range, leading to high costs. Moreover, as discussed in Sec.~\ref{sec:4D tensor}, the downlink signals from LEO satellites are affected by shadowing, which complicates the internal structure of the 3D RM. We conduct a singular value decomposition (SVD) operation on the top and bottom slices of the 3D RM and analyze the cumulative contribution of the principal components $ T_i=\sum_{k=1}^i \sigma_k / \sum_{k=1}^I \sigma_k$, where $\sigma_i$ is the $i$-th singular value of the slice. As illustrated in Fig.~\ref{fig:rank}, the top slice's first five principal components account for over 90$\%$ energy of the slice, indicating a low-rank characteristic, while the bottom slice's first five principal components contain less than 70$\%$, exhibiting a non-low-rank characteristic. This disparity in characteristics naturally poses challenges for 3D RM reconstruction. To investigate the performance of 3D RM reconstruction under sparse sensor constraints, we deploy two efficient tensor completion algorithms, Tmac~\cite{xu2013parallel} and KBR-TC~\cite{xie2017kronecker}. Fig.~\ref{fig:mse} presents the mean square error (MSE) between the reconstructed 3D RM and the ground-truth 3D RM at diverse sampling rates. It is evident that when the number of sensors is extremely sparse (missing rate exceeds 96$\%$), the overall reconstruction performance significantly deteriorates.}

\section{Framework Design}\label{sec:design}
\subsection{Problem Formulation}\label{sec:pf}

\needrev{The reconstruction of both 3D RM and CS are similar, as both aim to estimate unknown data from few samples. Evaluating the estimation error allows us to rigorously assess the performance of algorithms and models.}

In recent years, advances in supervised artificial intelligence have sparked widespread applications of NNs, further driving innovations in wireless communication~\cite{perenda2021learning,perenda2023contrastive,scalingi2024det,lin2024split,feng2023dynamic,sanchez2022airnn,zhang2021signal,lin2024efficient}. Simultaneously, unsupervised learning methods based on NNs have also seen rapid development~\cite{qayyum2022untrained}. Through hierarchical structures and nonlinear activation functions, these methods can effectively capture nonlinear features in data, accurately modeling the complex structures of real-world data, thus achieving high-precision estimation of unknown data from limited samples~\cite{luo2023low}.
Theoretical analysis in \cite{fan2021multi} shows that, compared to traditional linear optimization methods, unsupervised optimization methods based on NNs have a lower bound on estimation errors, especially in scenarios where a significant amount of necessary samples are missing. 

\begin{figure}[t]
\centering  
\subfigure[Cumulative contribution of the principal components v.s.  singular value]{
\label{fig:rank}
\includegraphics[width=0.24\textwidth]{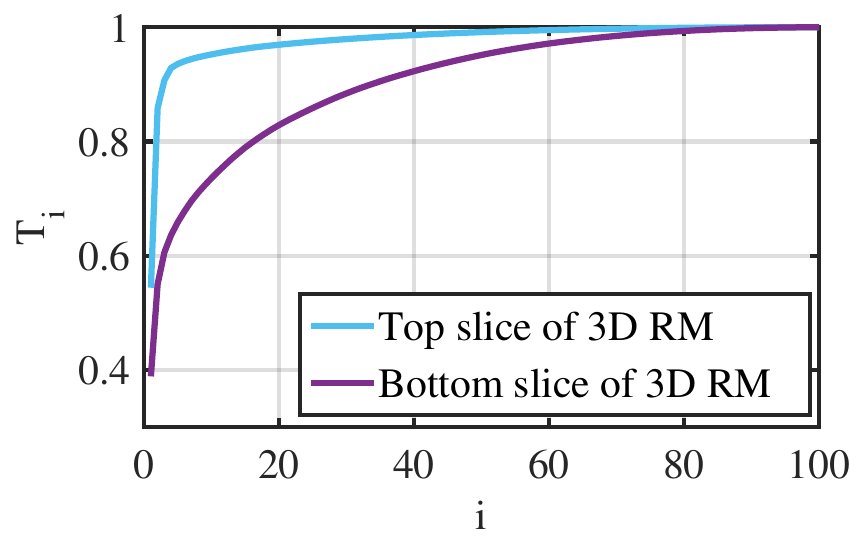}}
\subfigure[MSE v.s. Missing rate]{
\label{fig:mse}
\includegraphics[width=0.22\textwidth]{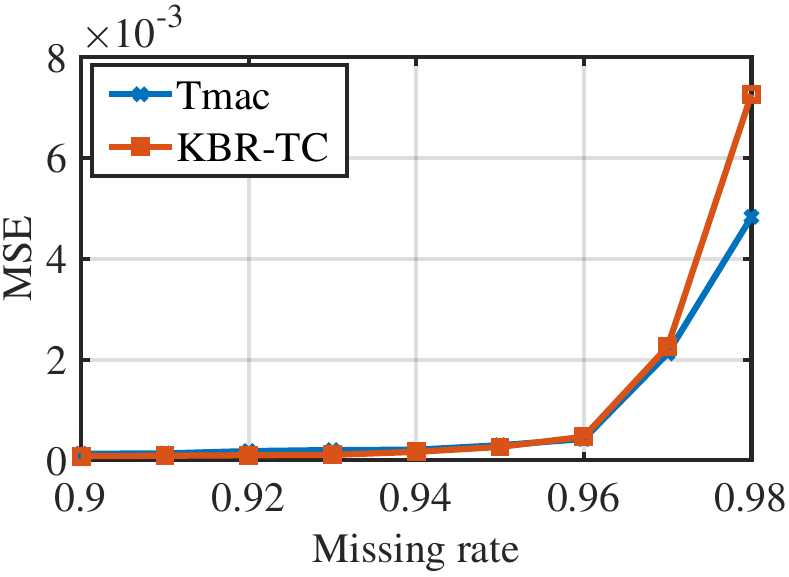}}
\caption{The shadowing effect on 3D RM (a) and reconstruction performance using Tmac~\cite{xu2013parallel} (a) and Tmac~\cite{xu2013parallel} (b) under diverse missing rates.}
\label{Fig.challenge1}
\end{figure}

Given the robust nonlinear modeling capabilities of NNs \needrev{in unsupervised learning}, an intuitive approach to construct 4D RM with limited samples is to employ an end-to-end NNs, which takes sub-Nyquist samples as input and unifies the CS and 3D RM reconstruction problems into a single loss function optimization, \needrev{transforming the optimization variables into the parameters} $\theta$ of the NNs. The formulation is as follows:
\begin{equation}
\mathcal{L}_{\text {Total}}(\theta)=\mathcal{L}_{\text {CS }}(\theta)+\mathcal{L}_{\text {Recon}}(\theta),
\label{eq:demo_loss}
\end{equation}
where $\mathcal{L}_{\mathrm{CS}}$ and $\mathcal{L}_{\mathrm{Recon}}$ correspond to the loss functions of CS and 3D RM, respectively, and $\mathcal{L}_{\mathrm{Total}}$ represents the overall loss function to be optimized. However, this approach faces two implementation challenges: first, \needrev{the optimization variable sizes for CS and 3D RM reconstruction differ}, making it difficult to coordinate a unified NN architecture. Second, simultaneously optimizing CS and 3D RM reconstruction tasks necessitates a complex NN structure and intricate loss functions, thereby complicating the achievement of stable convergence.


\begin{figure*}[t]
    \centering
    \includegraphics[width=1\linewidth]{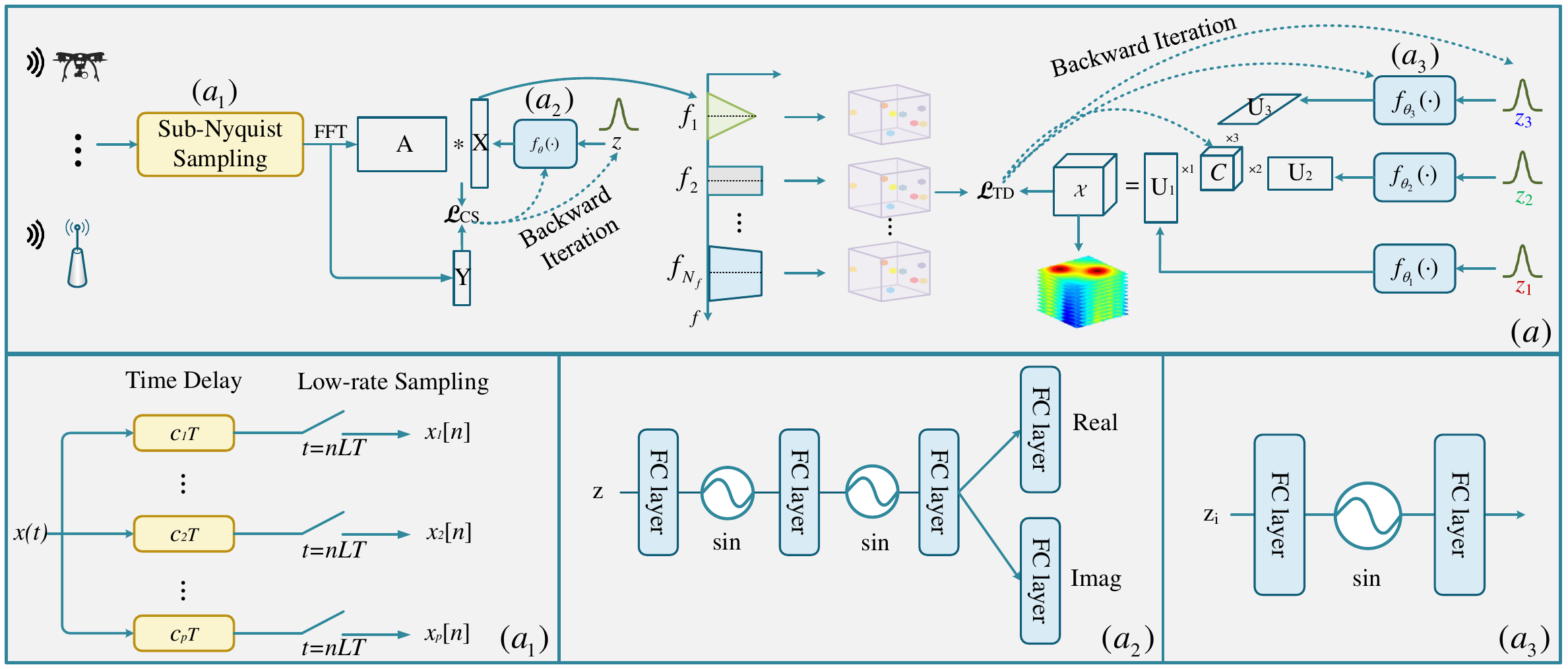}
    \caption{The overview of DeepRM framework ($a$); Sub-Nyquist sampling module ($a_1$); The NN architecture of $f_{\theta}(\cdot)$ ($a_2$) and $f_{\theta_i}(\cdot)$ ($i=1, 2,..., 3$) ($a_3$). }
    \label{fig:framework}
\end{figure*}

To address these issues, we convert the end-to-end method into a NNs-based framework named \name, splitting the Eqn.~\eqref{eq:demo_loss} into two sub-problems for separate optimization. As shown in Fig.~\ref{fig:framework}(a), the first part of \name, neural CS, solves the CS problem by minimizing the loss function $\mathcal{L}_{\mathrm{CS}}$, thereby reconstructing the power spectrum and obtaining the RSS information for each frequency, corresponding to step 1 in Fig.~\ref{fig:teaser}. The second part, neural TD, solves the 3D RM reconstruction problem by minimizing the loss function $\mathcal{L}_{\mathrm{TD}}$, thus obtaining the RSS information for each position in the 3D space, corresponding to step 2 in Fig.~\ref{fig:teaser}. The following sections will detail the construction of the loss functions and their optimization.
\subsection{Neural Compressive Sensing}\label{sec:NN_CS} 

As discussed in Sec.\ref{sec:sub-Nyquist} 2, due to hardware limitations, the sensors are constrained to sample wideband signals at sub-Nyquist rates. Multi-coset sampling (MCS)~\cite{song2019real,yang2021adaptive}, a well-established sub-Nyquist sampling technique, employs multiple parallel branches, each incorporating a delay unit and a low sampling rate ADC. This approach achieves compressed sampling through various combinations of delays and reduced sampling rates. Given its feasibility and comprehensible implementation, MCS is widely chosen for low-cost wideband signals sampling.

As illustrated in Fig.~\ref{fig:framework} ($a_1$), $x(t)$ is a band-limited original signal within the range $\left[ {0, B} \right]$, where $B$ represents the maximum frequency of the band-limited signal. The Nyquist sampling interval is denoted by $T=\frac{1}{2B}$. In this MCS approach, the signals are sampled in parallel by $P$ ADCs each with a sampling interval of $LT$ ($P < L$),  where $L$ is a constant representing the ratio of the ADC sampling interval to the Nyquist sampling interval. Each ADC is configured with a unique time delay. Consequently, the sampled sequence of the $j$-th ADC is given by
\begin{equation}
{x_j}\left[ n \right] = x\left( {nLT + {c_j}T} \right),\quad n = 1,2,...,N,
\end{equation}
where $c_j$ denotes a non-negative integer satisfying $0 \le {c_j} < L$ and ${c_j} \ne {c_k}\left( {\forall k \ne j} \right)$. $N$ represents the number of sampled data for each ADC. By employing FFT, multi-coset sampler follows this specific measurement procedure:
\begin{equation}
\mathbf{Y}=\mathbf{A} \mathbf{X}+\mathbf{N},
\label{eq:cs}
\end{equation}
where $\mathbf{Y} \in \mathbb{C}^{P \times N}$ is the measurement with
\begin{equation}
Y_{r, n}=L T e^{-j 2 \pi c_r n /(L N)} \sum_{i=1}^N x_r[i] e^{-j 2 \pi n i / N},
\end{equation}
and $\mathbf{A} \in \mathbb{C}^{P \times L}$ is the sensing matrix with
\begin{equation}
A_{r, l}=e^{j 2 \pi c_p \frac{(l-1)}{L}},
\end{equation}
and $\mathbf{X} \in \mathbb{C}^{L \times N}$ corresponds to the $L N$-points discrete Fourier transform of the discrete version of $x(t)$ by
\begin{equation}
X_{l, n}=\sum_{\tau=1}^{L N} x\left(\tau {T}\right) e^{-j 2 \pi \frac{(l-1) N+n}{L N}},
\end{equation}
and $\mathbf{N}$ is the additive noise caused by multiple factors such as channel noise and internal thermal noise of the receiver devices. \needrev{In a wideband spectrum, with narrow band signals occupying limited sub-bands, $\mathbf{X}$ have the row-sparse property that only a small number of rows (sub-band spectral) are non-zeros.}

The goal of CS is to recover $\mathbf{X}$ by solving the underdetermined Eqn.~\eqref{eq:cs}. Due to the sensing matrix $\mathbf{A}$ satisfying the restricted isometric property (RIP) and the block-sparse property of the wideband signal spectrum, $\mathbf{X}$ can be recovered by solving the following convex optimization problem~\cite{song2019real}:
\begin{equation}
\underset{\mathbf{X} \in \mathbb{C}^{L \times N}}{\arg \min }\|\mathbf{X}\|_{1} \text { s.t. }\|\mathbf{Y}-\mathbf{A} \mathbf{X}\|_{\mathbf{F}}<\epsilon,
\label{eq:cs_rec}
\end{equation}
where $\|\cdot\|_{1}$ is  to denote the number of non-zero rows, $\|\cdot\|_{F}$ is the Frobenius norm, and $\epsilon$ represents the noise power. The unconstrained equivalent form of Eqn.~\eqref{eq:cs_rec} is:
\begin{equation}
\underset{\mathbf{X} \in \mathbb{C}^{L \times N}}{\arg \min }\|\mathbf{Y}-\mathbf{A} \mathbf{X}\|_{F} +\lambda\|\boldsymbol{\mathbf{X}}\|_1,
\end{equation}
where $\lambda$ is a regularization parameter used to induce sparsity. Instead of directly optimizing for the variable $\mathbf{X}$ in Eqn.~\eqref{eq:cs_rec}, we can represent the optimization variable $\mathbf{X} = f_\theta(z)$, where $f_\theta(\cdot)$ represents the NNs. The input to $f_\theta(\cdot)$ is a learnable variable $\mathbf{z}$, which is initialized as a random variable following a normal distribution $N(0,1)$. The structure of $f_\theta(\cdot)$ is shown in Fig.~\ref{fig:framework} ($a_2$), which includes five fully connected (FC) layers and two activation function. Specifically, the output of the $(i+1)$-$th$ layer is:
\begin{equation}
\mathbf{h}_{i+1}=\phi\left(\mathbf{h}_{i} \mathbf{W}^{\top}_i+\mathbf{b}_i\right),
\end{equation}
where $\mathbf{W}_i$ is a learnable weight matrix, $\mathbf{h}_{1} = \mathbf{z}$, $\mathbf{b}_i$ is a bias vector and $\phi(\cdot)$ is a nonlinear activation function. Since the optimization variable $\mathbf{X}$ is a complex matrix, the last layer of the NNs branches into real and imaginary parts and obtains the final output through the\texttt{ Complex(Real,Imag)} operation. In this way, the CS optimization problem can be transformed into a loss function optimization problem:
\begin{equation}
    \mathcal{L}_{\mathrm{CS}}=\left\|\mathbf{Y}-\mathbf{A} f_{\theta}(\mathbf{z})\right\|_F+\lambda\left\|f_\theta(\mathbf{z})\right\|_1,
\end{equation}
where $\lambda$ represents regularization parameter. Through iterative back-propagation using gradient descent, the solution for the optimization variable $\mathbf{X}$ can be found in the nonlinear space. We employ
the popularized gradient evaluation tool \texttt{autograd} in
Pytorch~\cite{shrestha2022deep} to carry out the computation of gradient $\nabla_\theta \mathcal{L}_{\mathrm{CS}}$ and $\nabla_{\mathbf{Z}} \mathcal{L}_{\mathrm{CS}}$. The parameters of $f_\theta(\cdot)$, denoted as $\theta$ and variable $\mathbf{z}$ updating at iterations $t = 1, 2, \ldots$ are given by:

\begin{equation}
\begin{split}
\theta^{t+1} &= \theta^t-\eta \bar{\nabla}_\theta \mathcal{L}_{\mathrm{CS}}\left(\theta^t, \mathbf{z}^t\right) \\
\mathbf{z}^{t+1} &= \mathbf{z}^t-\eta \bar{\nabla}_{\mathbf{z}} \mathcal{L}_{\mathrm{CS}}\left(\theta^t, \mathbf{z}^t\right),
\end{split}
\end{equation}
where $\eta$ represents the learning rate, used to balance the convergence speed and stability. $\bar{\nabla}_\theta \mathcal{L}_{\mathrm{CS}}$ and $\bar{\nabla}_\mathbf{z} \mathcal{L}_{\mathrm{CS}}$ are gradient-based directions used to direct parameter updates, calculated by the \texttt{Adam} optimizer~\cite{shrestha2022deep} based on $\nabla_\theta \mathcal{L}_{\mathrm{CS}}$ and $\nabla_{\mathbf{Z}} \mathcal{L}_{\mathrm{CS}}$. The detailed procedure of neural CS is illustrated in Algorithm 1.

To accurately reconstruct the sparse signal $\mathbf{X}$, the model $f_{\theta}(\cdot)$ must adapt to the data structure while enforcing sparsity constraints. To balance sparsity and reconstruction performance, we propose a two-stage optimization approach, detailed in Algorithm 1. During the first stage, we jointly minimize reconstruction and sparsity loss to initialize the model and input variables. The second stage further optimizes these parameters, focusing solely on minimizing reconstruction loss for improved performance.

\begin{algorithm}
\caption{Neural Compressive Sensing Algorithm}
\begin{algorithmic}[1]
\STATE \textbf{Input:} Sensing matrix $\mathbf{A}$, observed measurements $\mathbf{Y}$;
\STATE \textbf{Initialization::} Initialize $f_\theta^1(\cdot)$ with activation function $\phi(\cdot)=\sin \left(\omega_0 \cdot\right)$, input variable $\mathbf{z}^1 \sim N(0,1)$, regularization parameter $\lambda$ and learning rate $\eta$;

\STATE \textbf{Joint Optimization Stage:}
\FOR{$t$ $= 1$ to $T$}
    \STATE Predict $\mathbf{X}_{\text{pred}} = f_{\theta}^t(\mathbf{z}^t)$;
    \STATE Compute CS loss: 
    
    $\mathcal{L}_{\text {CS }}=\left\|\mathbf{Y}-\mathbf{A} \mathbf{X}_{\text {pred }}\right\|_F+\lambda\left\|\mathbf{X}_{\text {pred }}\right\|_1$;
    \STATE Update $f_\theta^{t+1}(\cdot)$ and $\mathbf{z}^{t+1}$ via Eqn.(11);
\ENDFOR
\STATE Save the optimized $f_\theta^{T_{\text{joint}}}$ and $\mathbf{z}^{T_{\text{joint}}}$;

\STATE \textbf{Reconstruction Optimization Stage:}
\STATE Initialize by loading $f_\theta^{T_{\text{joint}}}$ and $\mathbf{z}^{T_{\text{joint}}}$;
\FOR{$t$ $= 1$ to $T_{\text{recon}}$}
    \STATE Predict $\mathbf{X}_{\text{pred}} = f_{\theta}^t(\mathbf{z}^1)$;
    \STATE Compute reconstruction loss: 
    
    $\mathcal{L}_{\mathrm{recon}}=\left\|\mathbf{Y}-\mathbf{A} \mathbf{X}_{\text {pred }}\right\|_F$;
    \STATE Backpropagation and update: 
    
    $\theta^{t+1}=\theta^t-\eta \bar{\nabla}_\theta \mathcal{L} \operatorname{recon}\left(\theta^t, \mathbf{z}^t\right)$, 
    
    $\mathbf{z}^{t+1}=\mathbf{z}^t-\eta \bar{\nabla}_{\mathrm{x}} \mathcal{L} \operatorname{recon}\left(\theta^t, \mathbf{z}^t\right)$;
\ENDFOR
\STATE \textbf{Output:} The recovered signal $\widehat{\mathbf{X}}=f_\theta^{T_{\text{recon}}}\left(\mathbf{z}^{T_{\text{recon}}}\right)$;

\end{algorithmic}
\end{algorithm}

\subsection{Neural Tensor Decomposition }\label{sec:NN_TD}

The reconstruction of 3D RM can be framed as a tensor completion problem, where efficient techniques like Canonical Polyadic (CP) decomposition\cite{li2023lightnestle}, Tucker decomposition\cite{fan2021multi}, and block term decomposition (BTD)\cite{zhang2020spectrum} can be applied. Tucker decomposition, with its flexibility, approximates a tensor through a core tensor and factor matrices, allowing for varying ranks across modes. This enables effective capture of multidimensional correlations, making Tucker decomposition particularly suitable for accurately reconstructing 3D RM.

For a 3D RM tensor (third-order tensor) $\mathbf{\mathcal { X }} \in \mathbb{R}^{I \times J \times K}$, assuming the rank of the Tucker decomposition is $(R_1, R_2, R_3)$, $\mathbf{\mathcal { X }}$ can be decomposed into a core tensor, denote as $\mathcal{G} \in \mathbb{R}^{R_1 \times R_2 \times R_3}$ and three factor matrices, denote as $\mathbf{U_1} \in \mathbb{R}^{I \times R_1}$, $\mathbf{U_2} \in \mathbb{R}^{J \times R_2}$ and $\mathbf{U_3} \in \mathbb{R}^{K \times R_3}$ respectively. The expression is as follows:
\begin{equation}
\mathbf{\mathcal { X }} \approx \mathcal{G} \times_1 \mathbf{U_1} \times_2 \mathbf{U_2} \times_3 \mathbf{U_3},
\end{equation}
where, the symbol $\times_k, k=1,2,3$ denotes the modal product between a tensor and a matrix. The expression of Tucker decomposition can also be written in matrix form:
\begin{equation}
\begin{split}
    \mathcal{X}_{(1)} &\approx \mathbf{U}_1 \mathcal{G}_{(1)} \underbrace{\left(\mathbf{U}_3 \otimes \mathbf{U}_2\right)^{\top}}_{\mathbf{V}_1}, \\
    \mathcal{X}_{(2)} &\approx \mathbf{U}_2 \mathcal{G}_{(2)} \underbrace{\left(\mathbf{U}_3 \otimes \mathbf{U}_1\right)^{\top}}_{\mathbf{V}_2}, \\
    \mathcal{X}_{(3)} &\approx \mathbf{U}_3 \mathcal{G}_{(3)} \underbrace{\left(\mathbf{U}_1 \otimes \mathbf{U}_2\right)^{\top}}_{\mathbf{V}_3},
\end{split}
\end{equation}
where, $\mathbf{\mathcal{X}}_{(1)}$, $\mathbf{\mathcal{X}}_{(2)}$, and $\mathbf{\mathcal{X}}_{(3)}$ are the matrices obtained by unfolding the tensor $\mathbf{\mathcal{X}}$ along the first, second, and third dimensions, respectively. Their sizes are $I \times (JK)$, $J \times (KI)$, and $K \times (IJ)$, respectively. $\mathcal{G}_{(i)}$ is the matrix obtained by unfolding the core tensor $\mathcal{G}$ along the first, second, and third dimensions, respectively. Their sizes are $R_1 \times (R_2 R_3)$, $R_2 \times (R_3 R_1)$, and $R_3 \times (R_1 R_2)$, respectively. The symbol $\otimes$ denotes the Kronecker product. 

The tensor $\mathbf{\mathcal{X}} \in \mathbb{R}^{I \times J \times K}$ contains only sparse measurements. The set $\Omega$ represents the index set of observed measurements. $\mathcal{P}_{\Omega}: \mathbb{R}^{I \times J \times K} \rightarrow \mathbb{R}^{I \times J \times K}$ denotes the orthogonal projection operating on the set $\Omega$, which is defined as:
\begin{equation}
\left[\mathcal{P}_{\Omega}(\mathcal{X})\right]_{i j k}= \begin{cases}x_{i j k}, & \text { if }(i, j, k) \in \Omega \\ 0, & \text { otherwise }\end{cases}
\end{equation}
Here, our goal is to learn the factor matrices $\mathbf{U}_i$ and core tensor $\mathcal{G}$ from the tensor $\mathbf{\mathcal { X }}$ by solving an optimization problem, thereby reconstructing the complete 3D RM. The optimization problem is formulated as follows:
\begin{align}
\operatorname*{arg\,min}_{\mathcal{G}, \mathbf{U_1}, \mathbf{U_2}, \mathbf{U_3}} &\frac{1}{2}\left\|\mathcal{P}_{\Omega}(\mathbf{\mathcal{X}})-\mathcal{P}_{\Omega}\left(\mathcal{G} \times_1 \mathbf{U_1} \times_2 \mathbf{U_2} \times_3 \mathbf{U_3}\right)\right\|_F^2 \notag \\
&+ \frac{\lambda}{2}\left(\|\boldsymbol{\mathcal{G}}\|_F^2+\|\mathbf{U_1}\|_F^2+\|\mathbf{U_2}\|_F^2+\|\mathbf{U_3}\|_F^2\right),
\label{eq:tensor_opt}
\end{align}
where $\lambda$ represents regularization parameter. This optimization problem can be solved using the alternating least squares (ALS):
\begin{equation}
\arg\min_{\mathbf{U}_i} \frac{1}{2}\left\|\mathcal{P}_{\Omega_i}\left(\mathcal{X}_{(i)}-\mathbf{U}_i \mathcal{G}_{(i)}\mathbf{V}_i\right)\right\|_F^2 + \frac{\lambda}{2}\left\|\mathbf{U}_i\right\|_F^2,
\end{equation}
where $\mathcal{P}_{\Omega_i}$ denotes $\mathcal{P}_{\Omega}$ unfolding along the $i$-th dimension, for $i = 1,2,3$. 

Similar to Sec.~\ref{sec:NN_CS}, we do not directly solve for the factor matrices $\mathbf{U_i}, i=1,2,3$, but instead convert the problem to solving for the NNs parameters and input variables $\left\{f_{\theta_i}(\cdot), \mathbf{z}_i\right\}_{i=1}^k$. The structure of $f_{\theta_i}(\cdot)$ with parameters $\theta_i$ is shown in Fig.~\ref{fig:framework} ($a_3$), which includes two FC layers and one activation function $\phi(\cdot)$. The optimization problem in Eqn.~\eqref{eq:tensor_opt} can be rewritten as loss function optimization problem:
\begin{equation}
\begin{aligned}
\mathcal{L}_{\mathrm{TD}} = & \left\|\mathcal{P}_{\Omega}(\mathcal{X}) 
- \mathcal{P}_{\Omega}\left(\mathcal{G} \times_1 f_{\theta_1}\left(\mathbf{z}_1\right) \right.\right. \\
& \left.\left. \times_2 f_{\theta_2}\left(\mathbf{z}_2\right) 
\times_3 f_{\theta_3}\left(\mathbf{z}_3\right)\right)\right\|_F^2,
\end{aligned}
\end{equation}
where $\mathcal{L}_{\mathrm{TD}}$ does not include a regularization term, as regularization is implemented through \texttt{weight decay} in Pytorch. By computing the gradients of NNs parameter $f_{\theta_i}(\cdot)$ and input variables $\mathbf{z}_i$, we can minimize the loss function to obtain the optimized factor matrices $\mathbf{\hat{U}}_i$ and core tensor $\mathcal{G}$. The overall process of neural TD is shown in Algorithm 2:

\begin{algorithm}
\caption{Neural Tensor Decomposition Algorithm}
\begin{algorithmic}[1]
\STATE \textbf{Input:} Incomplete tensor $\mathbf{\mathcal { X }}$;
\STATE \textbf{Initialization::} Initialize $f_{\theta_i}^1(\cdot)$ with activation function $\phi(\cdot)=\sin \left(\omega_0 \cdot\right)$, input variable $\mathbf{z}_i^1 \sim N(0,1)$, factor matrices $\mathbf{U}_i^1, i=1,2,3$, core tensor $\mathcal{G}^1$ regularization parameter $\lambda$ and learning rate $\eta$;
\FOR{$t$ $= 1$ to $T$}
    \FOR{$i$ $= 1$ to $3$}
    \STATE Calculate the gradient $\left\{\nabla_{\theta_i}^t \mathcal{L}_{\mathrm{TD}}, \nabla_{\mathbf{z}_i}^t \mathcal{L}_{\mathrm{TD}}\right\}_{i=1}^3$ via \texttt{autograd};
    \ENDFOR
    \STATE Calculate the gradient $\nabla_{\mathcal{G}}^t \mathcal{L}_{\mathrm{TD}}$ using \texttt{autograd};
    \STATE Update: 
    $\left\{\left\{\theta_i^{l+1}, \mathbf{Z}_i^{t+1}\right\}_{j=1}^3, \mathcal{G}^{t+1}\right\} \leftarrow \text{\texttt{Adam}} \text{ optimizer }\left(\left\{\nabla_{\theta_i}^{\prime} \mathcal{L}_{\mathrm{TD}}, \nabla_{\mathbf{Z}_i}^{\prime} \mathcal{L}_{\mathrm{TD}}\right\}_{i=1}^3, \nabla_g^{\prime} \mathcal{L}_{\mathrm{TD}}\right)$;
    
\ENDFOR
\STATE \textbf{Output:} The reconstructed 3D RM $\widehat{\mathcal{X}}=\mathcal{G}^T \times_1 \mathbf{U}_1^T \times_2 \mathbf{U}_2^T \times_3 \mathbf{U}_3^T$;
\end{algorithmic}
\end{algorithm}

\section{Implementation and Experimental setup}\label{sec:implementation}
\subsection{Experimental Platform}
Our proposed \name consists of two main steps: first, reconstructing power spectrum from sub-Nyquist samples; second, reconstructing the 3D RM for specific frequency. Below, we detail the process of obtaining sub-Nyquist data and the simulation of 3D RM.

\begin{figure}[t]
    \centering
    \includegraphics[width=0.75\linewidth]{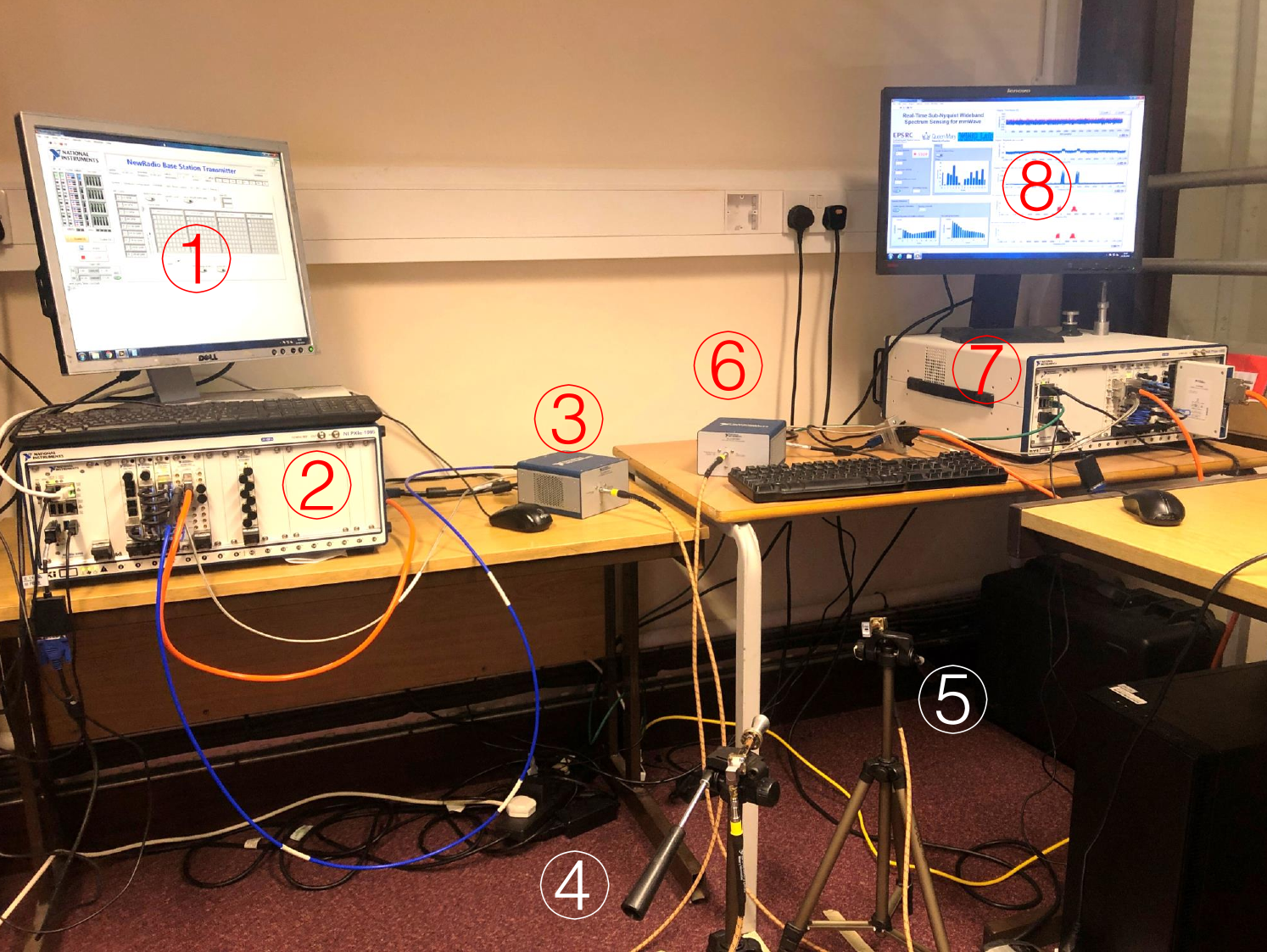}
    \caption{The implementation of NI SDR system. (1) (8) are Tx/Rx virtual panels, (2) (7) are Tx/Rx chassis, (3) (6) are mmWave Tx/Rx header and (4) (5) are Tx/Rx antennas.}
    \label{fig:NI}
\end{figure}

\textbf{Sub-Nyquist Platform:} Initially, we acquire Nyquist data using a software-defined radio (SDR) system from National Instruments (NI), which serves as both the transmitter (Tx) and receiver (Rx), as shown in Fig.~\ref{fig:NI}. These devices are equipped with modular, configurable hardware operating at a radio frequency centered at 28.5~\!GHz, within the Ka-band. The sensing bandwidth $B$ is 2~\!GHz. The baseband signal consists of in-phase (I) and quadrature (Q) components, covering a frequency range of -1~\!GHz to 1~\!GHz. At the receiver, this baseband signal is sampled by a single Nyquist ADC at a rate of 3.072~\!GHz. Next, we use the MCS ramework to compress the Nyquist data. The compression sampling factor $L$ is set to 40, with the number of low-speed ADCs, denoted as $P$ being either 10 or 16, and each ADC capturing 16 sampling points.



\begin{figure}[t]
    \centering
    \subfigure[Narrow beam coverage and corresponding RSS.]{
        \label{fig:motiv:cs:ss}
        \includegraphics[width=0.26\textwidth]{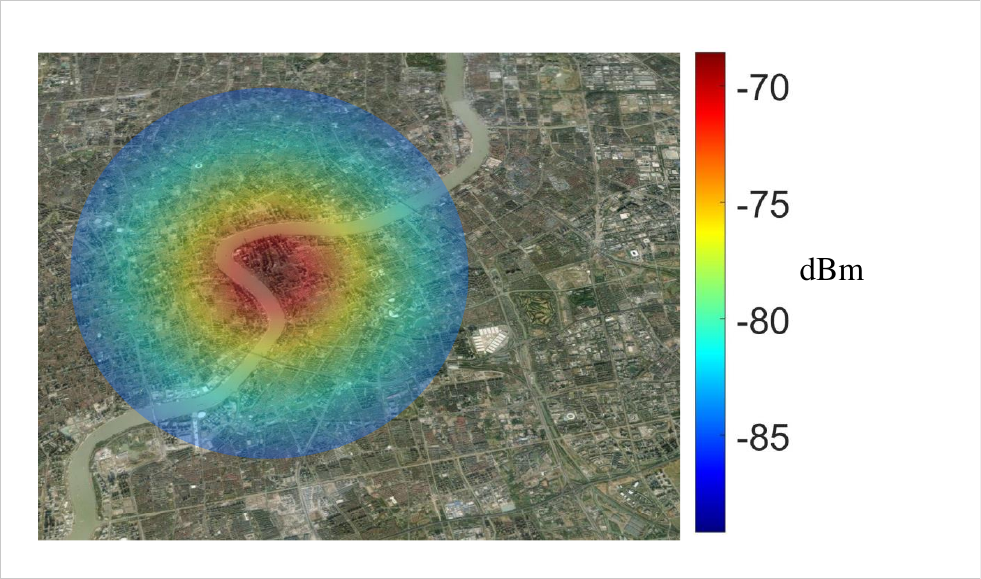}
    }
    \subfigure[Shadowing effect]{
        \label{fig:motiv:cs:rec}
        \includegraphics[width=0.18\textwidth]{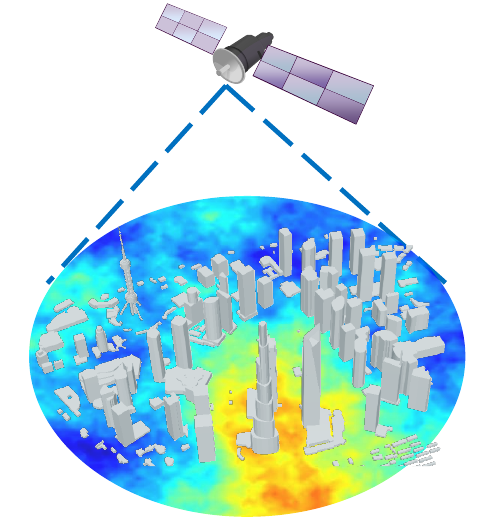}
    }
    \caption{Illustration of narrow beam coverage and  corresponding RSS (a) and shadowing effect in urban regions }
    \label{fig:3D RM}
\end{figure}

\textbf{3D RM Platform:} Given the lack of 3D RM simulation platforms for the scenario of LEO satellite networks, this paper fills this gap by developing a platform based on 3GPP technical reports~\cite{3GPP2020}. We set the sensing area to include a spatial range of 25~\!km by 25~\!km in the horizontal plane and 120~\!m in the vertical dimension, with each point in the 3D RM having a resolution of 250 m × 250 m × 10 m, resulting in a 3D RM size of 100 × 100 × 12. The number of narrow beams from LEO satellites covering the sensing area is set to 2. In our experiment, the orbital altitude of the LEO satellite is set to 550~\!km. Each narrow beam covers an area of approximately 160~\!km\(^2\), with a beam center RSS of around -70~\!dBm, which degrades following the free-space path loss. So the RSS distribution corresponding to the $r$-th narrow beam, denoted as $\boldsymbol{S}_r(\boldsymbol{y})$ can be expressed as
$\boldsymbol{S}_r(\boldsymbol{y})=\left\|\boldsymbol{y}-\boldsymbol{m}_r\right\|_2^{-\gamma_r}$,
in which $\boldsymbol{m}_r$ and $\gamma_r$ represents the location and path loss coefficient of the $r$-th narrow beam respectively and $\boldsymbol{y}=(i, j)$ represents the spatial coordinates. A visual representation of the narrow beam coverage is shown in Fig.~\ref{fig:3D RM}(a). 

Specifically, given that the narrow beam coverage area may encompass urban regions with dense buildings, the downlink signal of LEO satellite is subject to shadowing effect, as depicted in Fig.~\ref{fig:3D RM}(b). Hence, we incorporate a shadowing component~\cite{goldsmith2005wireless} into the RSS distribution in addition to the free-space path loss, so the RSS distribution corresponding to the $r$-th narrow beam can be rewritten as:
\begin{equation}
\boldsymbol{S}_r(\boldsymbol{y})=\left\|\boldsymbol{y}-\boldsymbol{r}_r\right\|_2^{-\gamma_r} 10^{v_r(\boldsymbol{y}) / 10},
\end{equation}
where $v_r(\boldsymbol{y})$ represents the correlated log-normal shadowing component, which is drawn from a zero-mean Gaussian distribution with a variance of $\sigma_r$. The auto-correlation between $\boldsymbol{y}$ and $\boldsymbol{y}^{\prime}$ adheres to $\mathbb{E}\left\{v_r(\boldsymbol{y}) v_r\left(\boldsymbol{y}^{\prime}\right)\right\}=\sigma_r^2 \exp \left(-\left\|\boldsymbol{y}-\boldsymbol{y}^{\prime}\right\|_2 / X_c\right)$, in which $X_c$ is referred to as the decorrelation distance. By adjusting $\sigma_r$ and $X_c$ to add varying degrees of shadowing at different vertical heights in the RM, we formed a 3D RM. The intensity of shadowing increases as the vertical height decreases.


\subsection{Experimental Setting}
This section primarily details the hyperparameter settings of \name, the baselines used in the experiments, and the evaluation metrics.

\textbf{Hyperparameter:} In the neural CS and neural TD algorithms, the parameter $\omega_0$ of the activation function $\phi(\cdot)$ is set to 10 and 2, respectively. The optimizable input variable $z$ in neural CS has a size of $L \times 1$, while the input variables ${z_1, z_2, z_3}$ in neural TD have sizes ${I \times 1, J \times 1, K \times 1}$ respectively. The optimizable core tensor $\mathcal{G}$ in neural TD is sized 25 × 25 × 12. The regularization parameter $\lambda$ in the loss function $\mathcal{L}_{\mathrm{CS}}$ is set to 1. 

\textbf{Baselines:} The baselines for comparison in power spectrum reconstruction stage are four classic compressive sensing algorithms: the SOMP and JB-HTP~\cite{song2019real} algorithms based on greedy search, the FCPSE algorithm~\cite{yang2019fast} based on compressive covariance and PCSBL-GAMP~\cite{fang2016two} algorithm based on sparse bayesian learning. Due to the significantly larger 3D RM size in the LEO satellite scenario compared to the TN scenario, we do not choose the method used for TN scenario ~\cite{shen20213d,wang2023sparse} as the baseline for comparison in 3D RM reconstruction. The baselines for 3D RM reconstruction are Kriging~\cite{setianto2013comparison} interpolation and prevalent tensor completion algorithms: Tmac~\cite{xu2013parallel}, KBR-TC\cite{xie2017kronecker}, and LNOP~\cite{chen2020robust}.

\textbf{Evaluation Metrics:} For the power spectrum reconstruction stage, we use MSE and the receiver operating characteristic (ROC) curve of spectrum sensing as evaluation metrics. For the 3D RM reconstruction stage, we use peak signal-to-noise ratio (PSNR) and structural similarity (SSIM) as evaluation metrics. Higher values of PSNR and SSIM signify better algorithmic performance.
\section{Evaluation}\label{sec:evaluation}
This section evaluates the performance of DeepRM in both the power spectrum reconstruction and 3D RM reconstruction stages.
\subsection{Power Spectrum Reconstruction}
\begin{figure}[t]
\centering  
\subfigure[Ground-truth]{
\label{fig:true spectrum}
\includegraphics[width=0.15\textwidth]{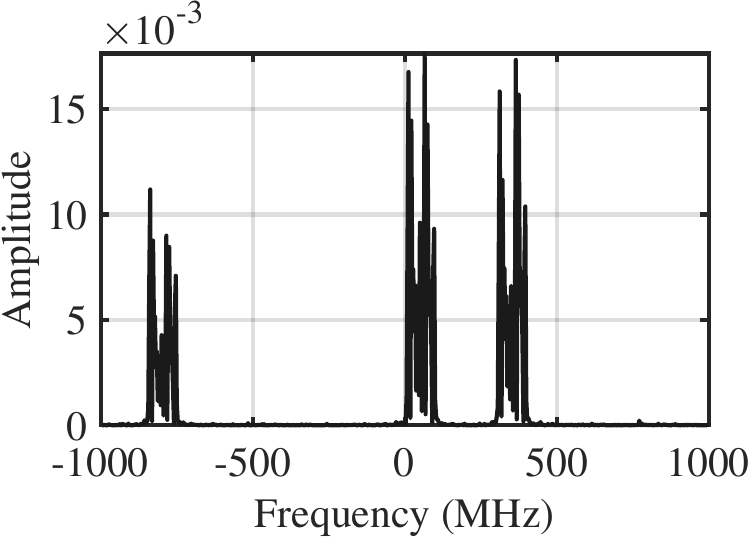}}
\subfigure[FCPSE]{
\label{fig:fcpse}
\includegraphics[width=0.15\textwidth]{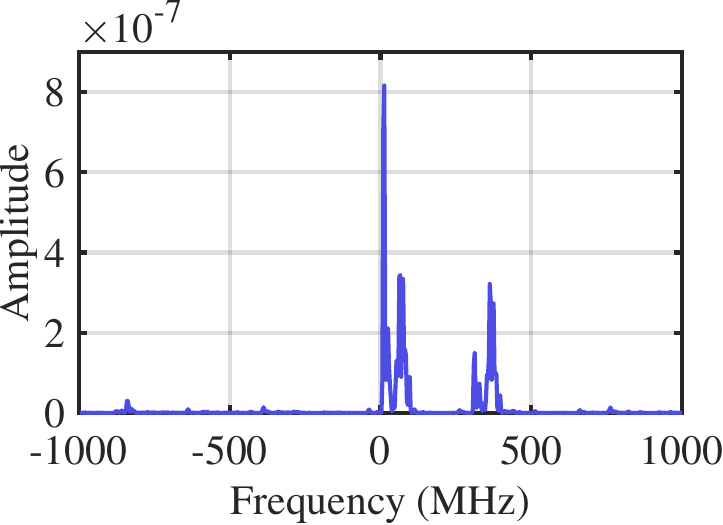}}
\subfigure[JB-HTP]{
\label{fig:somp}
\includegraphics[width=0.15\textwidth]{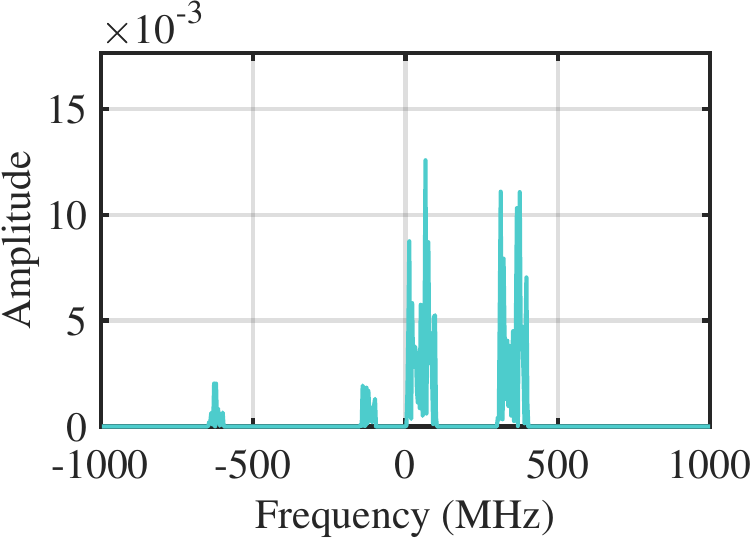}}\\
\subfigure[SOMP]{
\label{fig:jb-htp}
\includegraphics[width=0.15\textwidth]{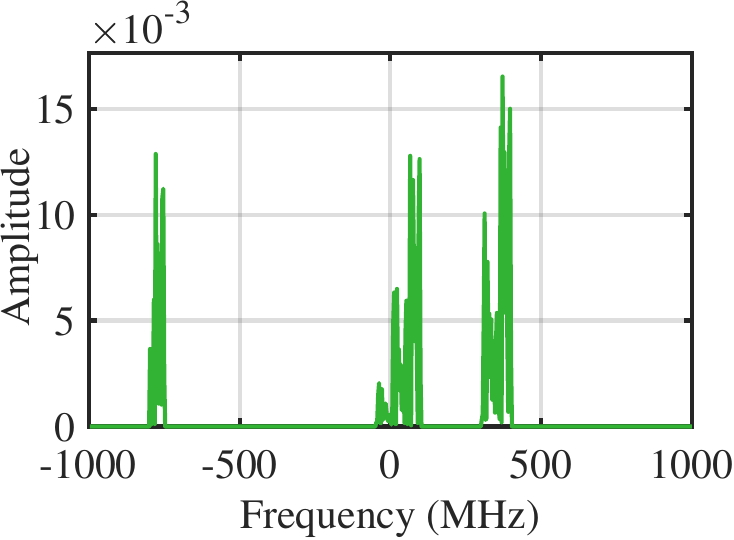}}
\subfigure[PCSBL-GAMP]{
\label{fig:sbl}
\includegraphics[width=0.15\textwidth]{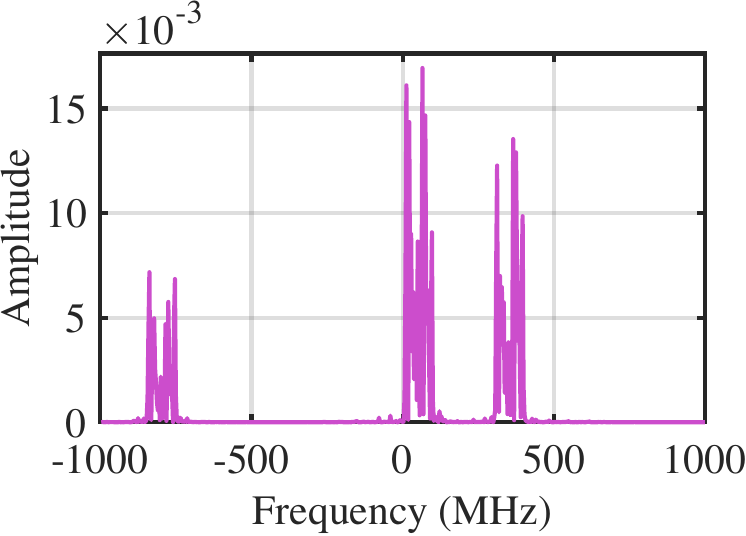}}
\subfigure[DeepRM]{
\label{fig:ncs}
\includegraphics[width=0.15\textwidth]{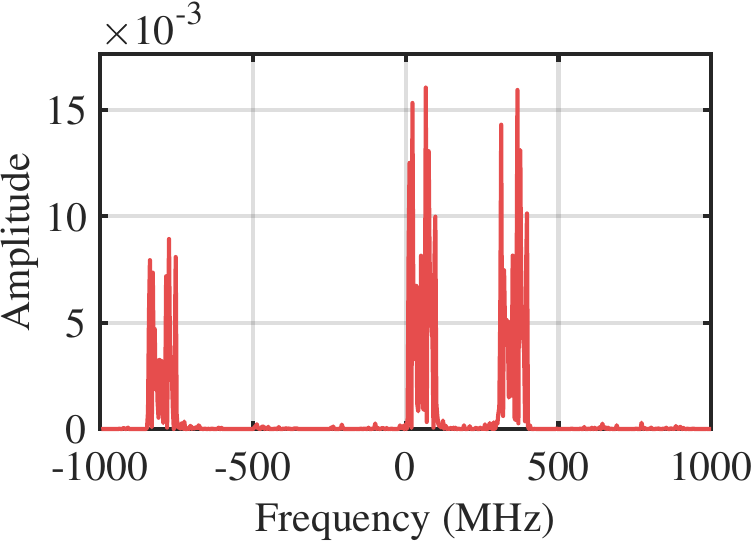}}
\caption{The comparison of reconstructed power spectrum for DeepRM and FCPSE~\cite{yang2019fast} (b), JB-HTP (c), SOMP (d)~\cite{song2019real}, PCSBL-GAMP~\cite{fang2016two} with $P=16$.}
\label{fig:16cosets}
\end{figure}

Fig.~\ref{fig:16cosets} illustrates the comparison between the reconstructed power spectrum obtained by DeepRM and various baseline algorithms with the true power spectrum within the MCS framework, where the number of cosets $P$~(a.k.a. number of low-speed ADCs) is set to 16, corresponding to a sampling rate of 2/5 of the Nyquist rate. When the narrow beams of LEO satellites occupy three frequency bands, the FCPSE and JB-HTP algorithms' reconstructed power spectra only reflect the RSS information of two bands, omitting one, and show significant amplitude deviations from the true power spectrum. In contrast, the SOMP, PCSBL-GAMP, and DeepRM algorithms accurately reflect the RSS information for all three bands, with DeepRM's reconstructed power spectrum amplitude being the closest to the true power spectrum.

\begin{figure}[t]
\centering  
\subfigure[Ground-truth]{
\label{fig:true spectrum}
\includegraphics[width=0.15\textwidth]{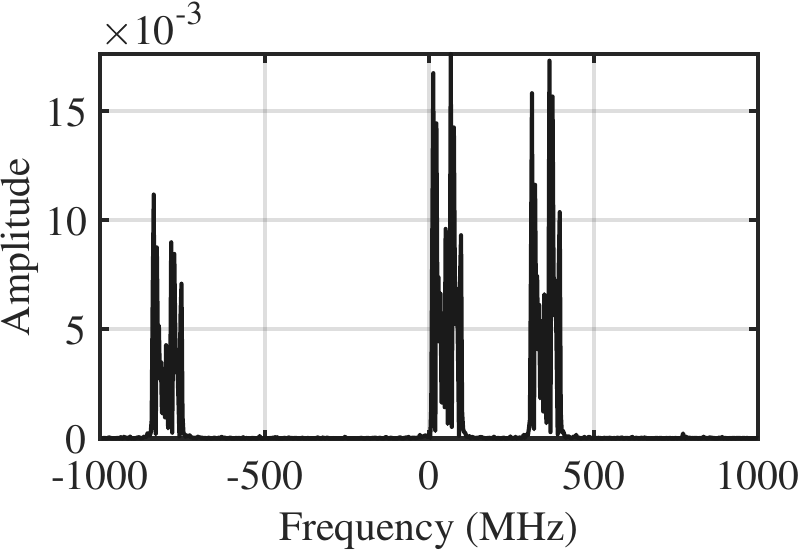}}
\subfigure[FCPSE]{
\label{fig:fcpse}
\includegraphics[width=0.15\textwidth]{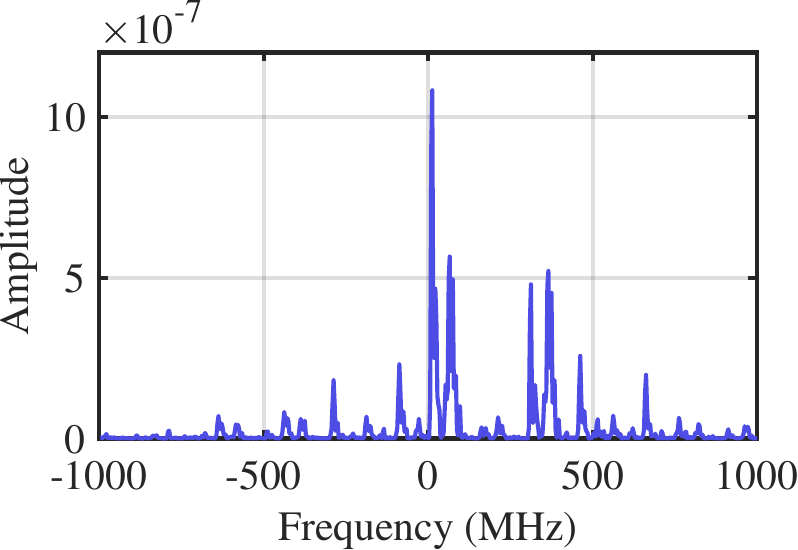}}
\subfigure[JB-HTP]{
\label{fig:somp}
\includegraphics[width=0.15\textwidth]{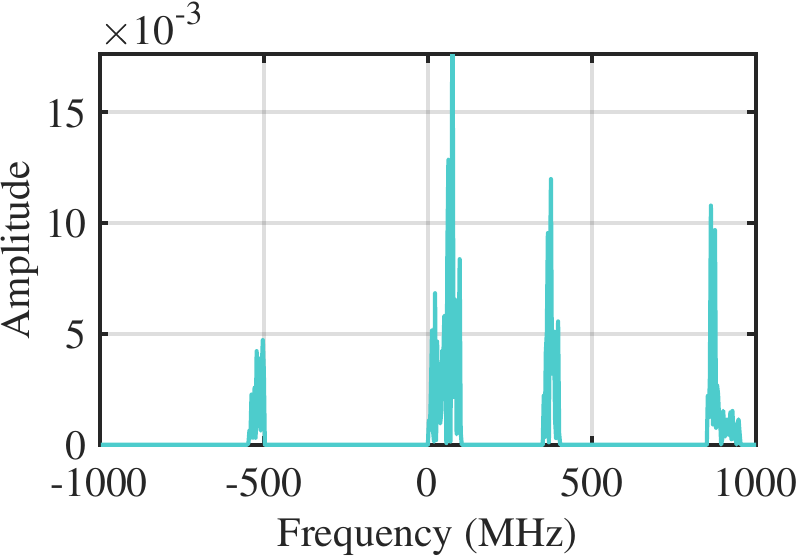}}\\
\subfigure[SOMP]{
\label{fig:jb-htp}
\includegraphics[width=0.15\textwidth]{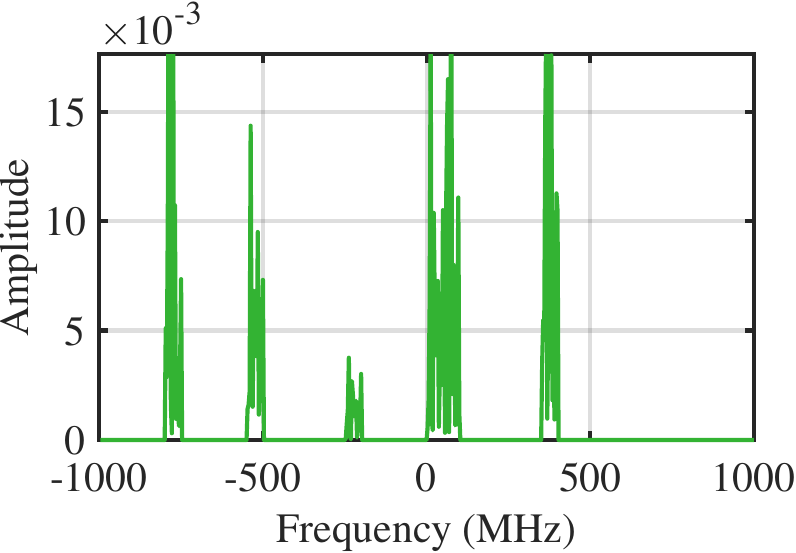}}
\subfigure[PCSBL-GAMP]{
\label{fig:sbl}
\includegraphics[width=0.15\textwidth]{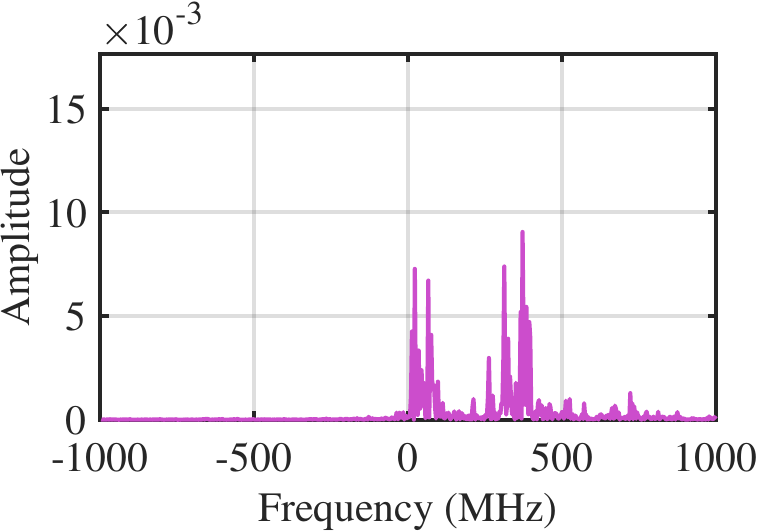}}
\subfigure[DeepRM]{
\label{fig:ncs}
\includegraphics[width=0.15\textwidth]{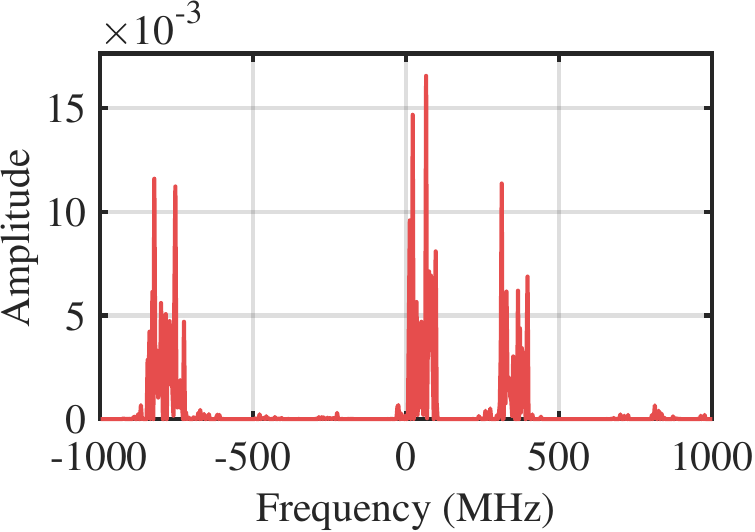}}
\caption{The comparison of reconstructed power spectrum for DeepRM and FCPSE~\cite{yang2019fast} (b), JB-HTP (c), SOMP (d)~\cite{song2019real}, PCSBL-GAMP~\cite{fang2016two} with $P=10$.}
\label{fig:10cosets}
\end{figure}
Fig.~\ref{fig:10cosets} presents the power spectrum reconstruction results with the number of cosets $P$ set to 10, corresponding to a sampling rate of 1/4 of the Nyquist rate. The performance of the baseline algorithms degrades significantly under these conditions, resulting in reconstructed power spectra that markedly deviate from the true power spectrum and fail to accurately identify the positions of the occupied bands. This indicates that these algorithms require a sufficient number of cosets to perform effectively, even under sub-Nyquist sampling. In contrast, DeepRM maintains high performance with a significantly reduced number of cosets, accurately reconstructing the RSS information of the occupied bands. This highlights potential and applicability of DeepRM in scenarios with hardware constraints.

\begin{figure}[b]
\centering  
\subfigure[P=16]{
\label{fig:roc_16}
\includegraphics[width=0.23\textwidth]{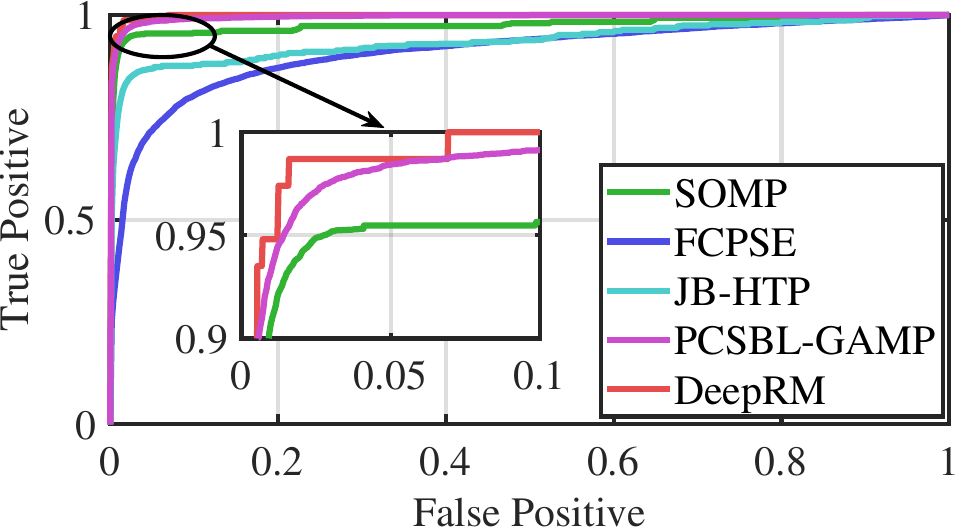}}
\subfigure[P=10]{
\label{fig:roc_10}
\includegraphics[width=0.23\textwidth]{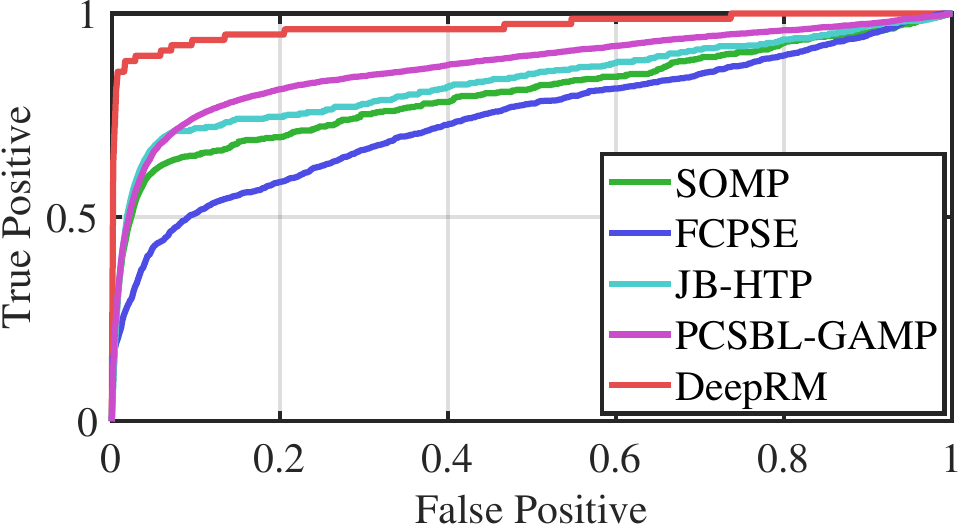}}
\caption{The comparison of spectrum sensing ROC curves for DeepRM and JB-HTP, SOMP~\cite{song2019real}, PCSBL-GAMP~\cite{fang2016two} with different $P$.}
\label{Fig.ROC}
\end{figure}

Fig.~\ref{Fig.ROC} presents the ROC curves for spectrum sensing, based on power spectrum reconstruction results from DeepRM and baseline algorithms over 1000 Monte Carlo experiments. The horizontal axis is the False Positive Rate (FPR) and the vertical axis is the True Positive Rate (TPR). Baseline algorithms perform well with many cosets but degrade significantly as the number of cosets decrease, indicating a strong reliance on the amount of sampled data. In contrast, DeepRM performs well with both high and low coset numbers. Even in hardware-constrained scenarios with fewer cosets, its ROC curve remains near the top-left corner, indicating high TPR even at low FPR. This demonstrates DeepRM's ability to accurately identify occupied frequency bands.


\begin{figure}[t]
    \centering
    \includegraphics[width=0.95\linewidth]{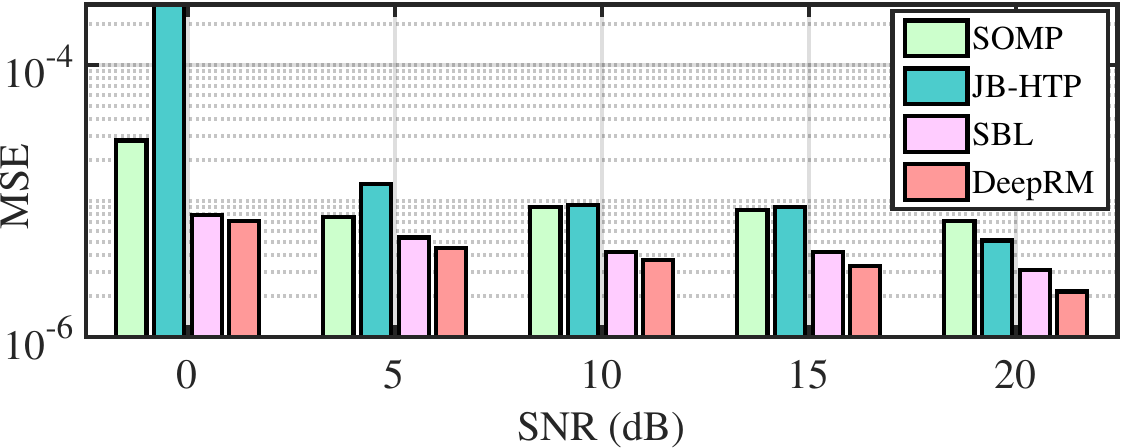}
    \caption{The comparison of the power spectrum reconstruction MSE for DeepRM and JB-HTP, SOMP~\cite{song2019real}, PCSBL-GAMP~\cite{fang2016two} with $P=10$.}
    \label{fig:MSE_SP}
\end{figure}

To further demonstrate the advantages of DeepRM in hardware-constrained scenarios, we conducted 1000 Monte Carlo experiments to compare the MSE between the reconstructed power spectrum by DeepRM and baseline algorithms with the true power spectrum under different signal-to-noise Ratio (SNR), as shown in Fig.~\ref{fig:MSE_SP}. It can be observed that DeepRM consistently achieves the lowest MSE across all SNR conditions, with particularly outstanding performance at low SNR, demonstrating its robustness and superiority in various noise environments.

\subsection{3D RM Reconstruction}

\begin{figure}[b]
    \centering
    \includegraphics[width=1\linewidth]{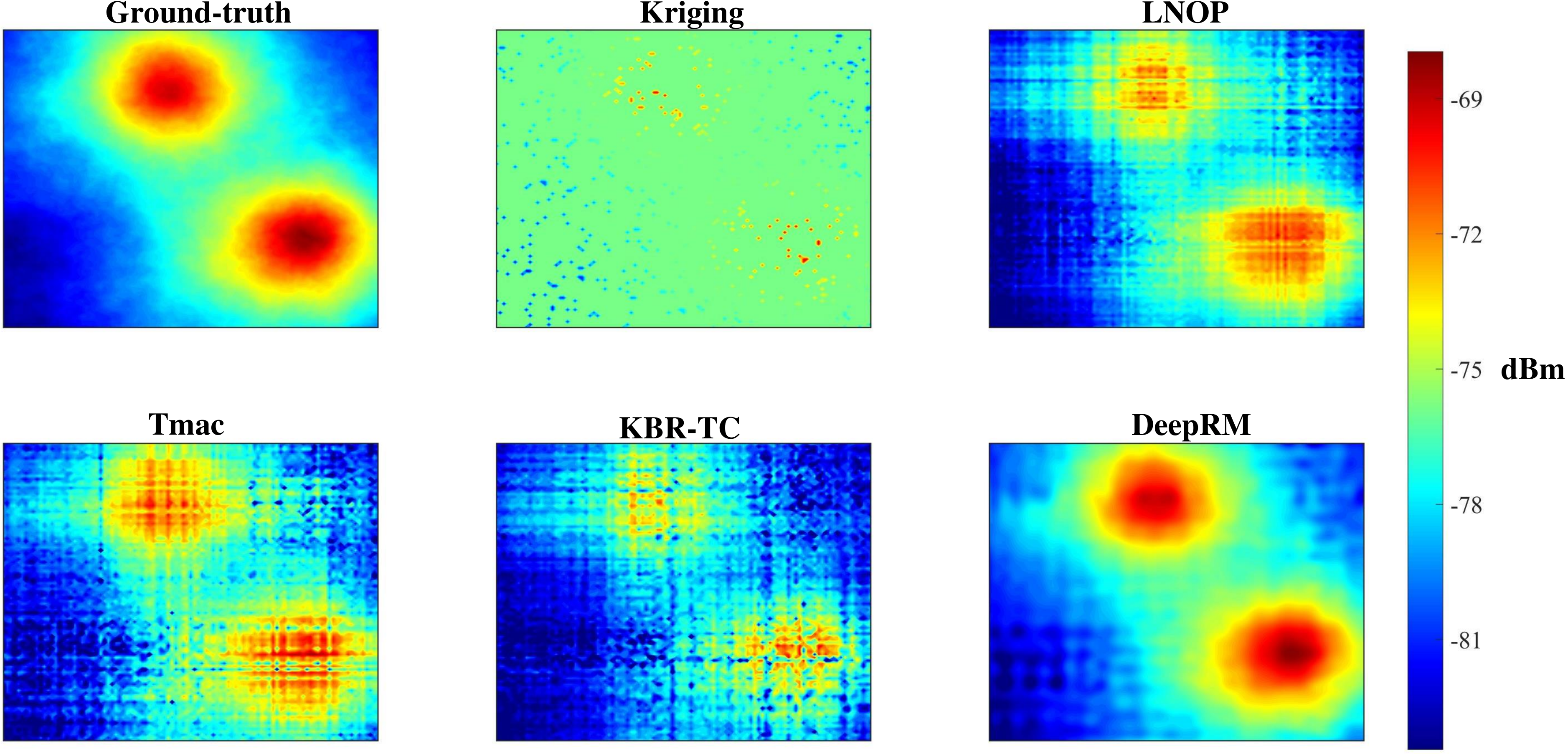}
    \caption{The 3D RM reconstruction performance of $2$-nd slice for \name and Kriging~\cite{setianto2013comparison}, Tmac~\cite{xu2013parallel}, KBR-TC\cite{xie2017kronecker}, and LNOP~\cite{chen2020robust}, where the missing rate is set to 0.98.}
    \label{fig:3D_RM1}
\end{figure}

Fig.~\ref{fig:3D_RM1} compares the visualizations of the 3D RM reconstructed by DeepRM and various baseline algorithms against the true 3D RM. The chosen visualization is the second slice of the 3D RM, minimally affected by shadowing, with a missing rate of 0.98, indicating an extremely sparse sensor deployment. It can be observed that Kriging interpolation nearly loses all RSS information, and KBR-TC struggles with accurate distribution. Although Tmac and KBR-TC recover the coverage areas of two narrow beams, their signal strength and distribution are inaccurate, with significant strip noise and loss of high-intensity RSS information. In contrast, DeepRM excels in recovering RSS information, showing a fine-grained overall distribution with minimal noise. Both the signal coverage area and high-intensity RSS information are well reconstructed, closely matching the true values, demonstrating DeepRM's accuracy and robustness in handling sparse data.

\begin{figure}[t]
    \centering
    \includegraphics[width=1\linewidth]{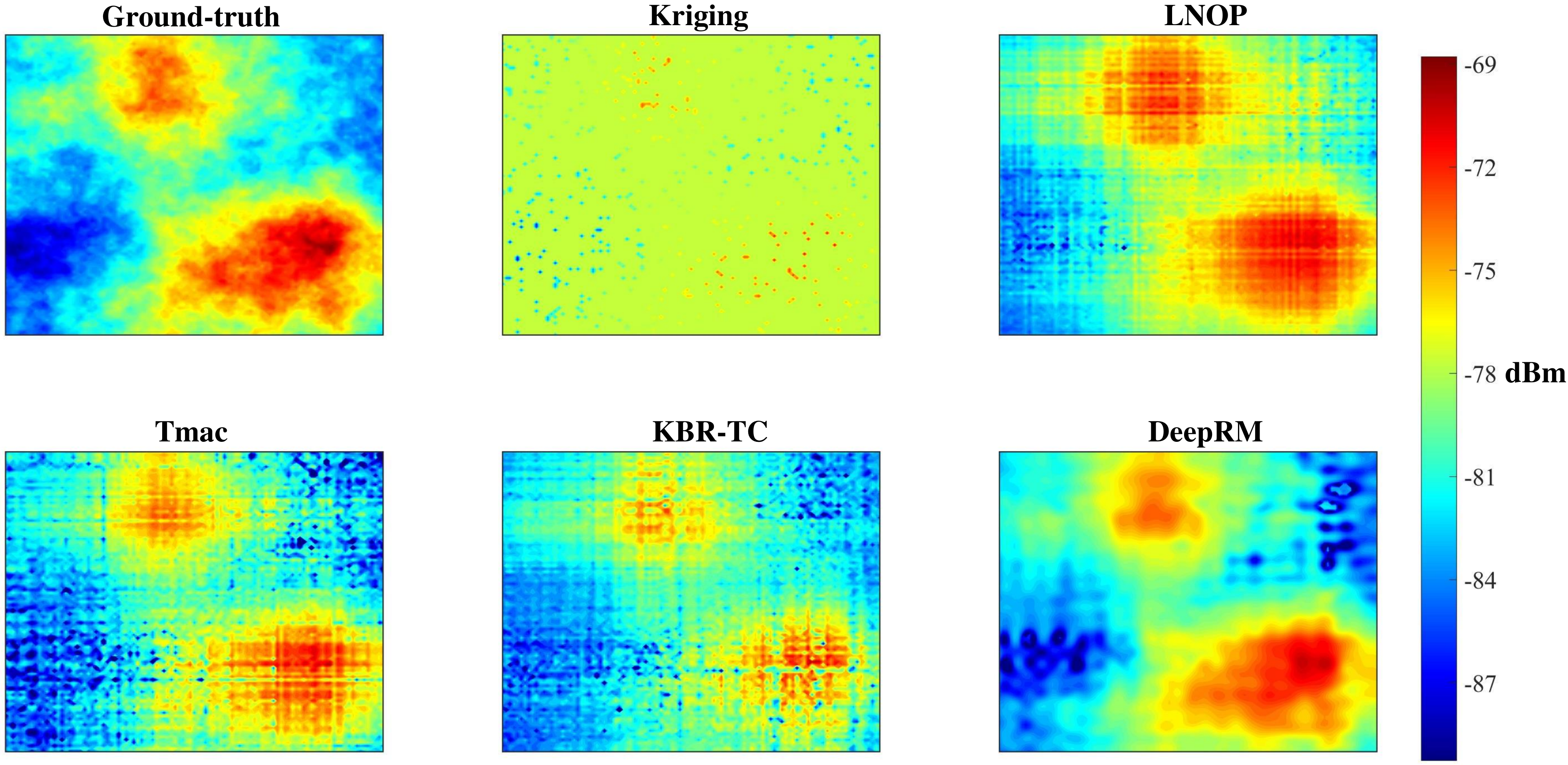}
    \caption{The 3D RM reconstruction performance of $12$-th slice for \name and Kriging~\cite{setianto2013comparison}, Tmac~\cite{xu2013parallel}, KBR-TC\cite{xie2017kronecker}, and LNOP~\cite{chen2020robust}, where the missing rate is set to 0.98.}
    \label{fig:3D_RM2}
\end{figure}



Fig.~\ref{fig:3D_RM2} shows the visualization comparison of 3D RM reconstruction when the signal is subjected to heavy shadowing, with the chosen visualization slice being the 12-th slice of the 3D RM. It can be observed that when the shadowing is strong, the RSS distribution becomes more complex, and the baseline methods struggle to reconstruct a clear RSS distribution. In contrast, DeepRM accurately reconstructs the fine-grained contours of the RSS distribution and the strong RSS information.

\begin{figure}[t]
\centering  
\subfigure[PSNR]{
\label{fig:motiv:cs:ss}
\includegraphics[width=0.23\textwidth]{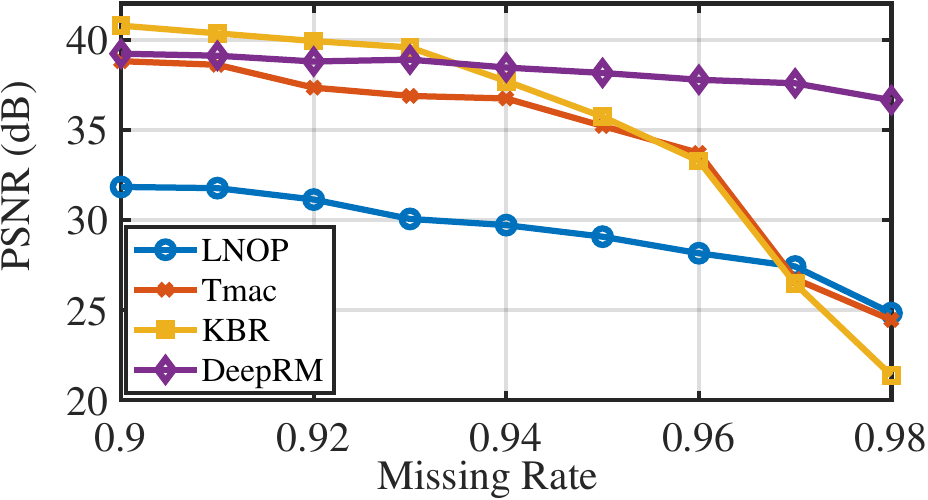}}
\subfigure[SSIM]{
\label{fig:motiv:cs:rec}
\includegraphics[width=0.23\textwidth]{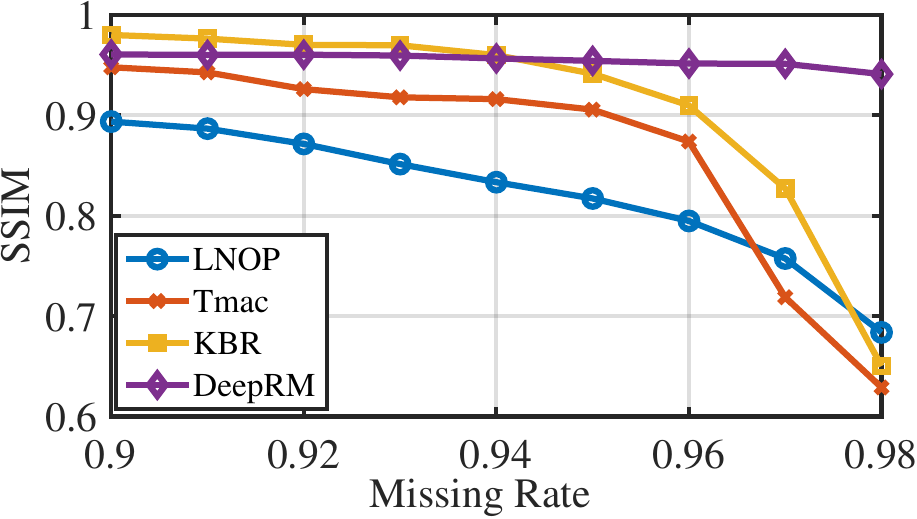}}
\caption{The comparison of \name and Tmac~\cite{xu2013parallel}, KBR-TC\cite{xie2017kronecker}, and LNOP~\cite{chen2020robust} in PSNR and SSIM at diverse missing rates.}
\label{fig:PSNR}
\end{figure}

Fig.~\ref{fig:PSNR} presents a comparison of PSNR and SSIM values for various reconstruction algorithms under different missing rates, excluding Kriging interpolation. It can be observed that at low missing rates, the Tmac, KBR-TC, and DeepRM achieve higher PSNR and SSIM values compared to the LNOP algorithm. However, as the missing rate increases, the performance of the Tmac and KBR-TC algorithms rapidly deteriorates, with their PSNR and SSIM values dropping sharply, indicating a significant decline in their reconstruction performance. In contrast, DeepRM remains stable, with only a slight decrease in PSNR and SSIM values, demonstrating its robust reconstruction capability even with limited samples. This highlights the potential and value of the DeepRM in sparse sensors condition.




\section{Related Work}\label{sec:related_work}
As a pivotal technique for acquiring 4D RM,  power spectrum reconstruction and RM reconstruction have emerged as prominent research hotspots in the academic community in recent years.

\textbf{Power Spectrum Reconstruction}: CS provides an effective solution for wideband spectrum sensing and reconstruction. The seminal work in~\cite{mishali2009blind} proposed an algorithm based on compressed power spectrum estimation, using a low-cost MCS receiver with the goal of reconstructing the wideband power spectrum instead of the original wideband signal. Subsequently, many studies have focused on the problem of compressed spectrum reconstruction under MCS. The work in~\cite{yang2021adaptive} designed a low-sampling-cost adaptive compressed spectrum sensing scheme based on MCS, significantly reducing sampling costs and computational complexity while maintaining the performance of traditional algorithms. The work in ~\cite{song2022approaching} conducted numerical simulations and verifications on the theoretical limits of spectrum sensing and reconstruction under the MCS framework. The work in~\cite{song2019real} combined random search and greedy pursuit strategies to design effective sampling patterns under different conditions, significantly improving the accuracy of spectrum sensing and reconstruction in dynamic environments.

\textbf{RM Reconstruction}: The work in~\cite{shen20213d} models 3D RM reconstruction as a 3D CS problem, solved using a 3D spatial subspace orthogonal matching pursuit algorithm. The work in~\cite{wang2023sparse} combines SBL with the maximum and minimum distance clustering algorithm for efficient 3D RM reconstruction. The work in~\cite{hu20233d} utilizes generative adversarial networks (GANs) for 3D RM reconstruction, which demonstrates good performance with limited samples. However, although it performs well with limited samples, it requires a large deployment of sensors, which significantly increases the costs. In the scenario of multi-frequency 2D RM, TD are
extensively used for some works~\cite{zhang2020spectrum,chen2022constrained}, which apply tensor TD to 3D RM reconstruction by decomposing the 3D RM tensor into power density spectrum (PSD) and spatial loss field (SLF) for separate estimation. With the rise of artificial intelligence, some DL-based methods have also garnered widespread attention in the academic community. The work in~\cite{teganya2021deep} utilize an Autoencoder (AE) to learn the spatial structure of propagation phenomena from a large dataset of measurements to achieve high-precision 3D RM reconstruction. The work in \cite{shrestha2022deep} leverages AE to fit the SLF of each emitter, thereby significantly reducing the offline training cost.

\section{Conclusion}\label{sec:conclusion}
In this paper, we propose a deep unsupervised neural networks-based framework  \name to construct the 4D RM of LEO satellite networks with limited samples. \name includes two main components: The first part is a neural compressive sensing module, which utilizes neural networks to nonlinearly represent and solve the compressive sensing problem, enabling low-error power spectrum reconstruction despite sampling rate constraints. The second part is a neural tensor decomposition module, which combines Tucker decomposition with neural networks to nonlinearly optimize factor matrices, allowing the reconstruction of a complete 3D RM for each frequency even with sparse sensors. Extensive experiments show that \name outperforms corresponding state-of-the-art baselines, especially with limited samples. These results demonstrate the effectiveness of DeepRM in accurately constructing 4D RM from sparse data. In the future, we would like to design an efficient resource scheduling strategy based on \name.  As a potential future direction, we are looking forward to extending our \name to improve the performance of various applications such as distributed learning systems~\cite{zhang2024fedac,lin2024adaptsfl,hu2024accelerating,leo2025split,zhang2024satfed,lin2024hierarchical,zhang2025lcfed} and large language models~\cite{lin2023pushing,hu2024agentscodriver,fang2024automated,hu2021lora,hu2024agentscomerge,lin2024splitlora}.

\bibliographystyle{IEEEtran}
\bibliography{main}

\begin{thebibliography}{10}
\providecommand{\url}[1]{#1}
\csname url@samestyle\endcsname
\providecommand{\newblock}{\relax}
\providecommand{\bibinfo}[2]{#2}
\providecommand{\BIBentrySTDinterwordspacing}{\spaceskip=0pt\relax}
\providecommand{\BIBentryALTinterwordstretchfactor}{4}
\providecommand{\BIBentryALTinterwordspacing}{\spaceskip=\fontdimen2\font plus
\BIBentryALTinterwordstretchfactor\fontdimen3\font minus \fontdimen4\font\relax}
\providecommand{\BIBforeignlanguage}[2]{{%
\expandafter\ifx\csname l@#1\endcsname\relax
\typeout{** WARNING: IEEEtran.bst: No hyphenation pattern has been}%
\typeout{** loaded for the language `#1'. Using the pattern for}%
\typeout{** the default language instead.}%
\else
\language=\csname l@#1\endcsname
\fi
#2}}
\providecommand{\BIBdecl}{\relax}
\BIBdecl

\bibitem{zhao2024leo}
Z.~Zhao, Z.~Chen, Z.~Lin, W.~Zhu, K.~Qiu, C.~You, and Y.~Gao, ``Leo satellite networks assisted geo-distributed data processing,'' \emph{IEEE Wireless Communications Letters}, 2024.

\bibitem{yuan2023graph}
H.~Yuan, Z.~Chen, Z.~Lin, J.~Peng, Z.~Fang, Y.~Zhong, Z.~Song, X.~Wang, and Y.~Gao, ``Graph learning for multi-satellite based spectrum sensing,'' in \emph{Proc. ICCT}, 2023, pp. 1112--1116.

\bibitem{michel2022first}
F.~Michel, M.~Trevisan, D.~Giordano, and O.~Bonaventure, ``{A first Look at Starlink Performance},'' in \emph{Proceedings of the 22nd ACM Internet Measurement Conference}, 2022, pp. 130--136.

\bibitem{lin2024fedsn}
Z.~Lin, Z.~Chen, Z.~Fang, X.~Chen, X.~Wang, and Y.~Gao, ``Fedsn: A federated learning framework over heterogeneous leo satellite networks,'' \emph{IEEE Transactions on Mobile Computing}, 2024.

\bibitem{yuan2024satsense}
H.~Yuan, Z.~Chen, Z.~Lin, J.~Peng, Z.~Fang, Y.~Zhong, Z.~Song, and Y.~Gao, ``Satsense: Multi-satellite collaborative framework for spectrum sensing,'' \emph{arXiv preprint arXiv:2405.15542}, 2024.

\bibitem{xiao2022leo}
Z.~Xiao, J.~Yang, T.~Mao, C.~Xu, R.~Zhang, Z.~Han, and X.-G. Xia, ``{LEO Satellite Access Network (LEO-SAN) Towards 6G: Challenges and Approaches},'' \emph{{IEEE} Wireless Commun}, 2022.

\bibitem{ma2023network}
S.~Ma, Y.~C. Chou, H.~Zhao, L.~Chen, X.~Ma, and J.~Liu, ``{Network Characteristics of LEO Satellite Constellations: A Starlink-Based Measurement from End Users},'' in \emph{Proc. INFOCOM}, 2023, pp. 1--10.

\bibitem{liu2018space}
J.~Liu, Y.~Shi, Z.~M. Fadlullah, and N.~Kato, ``{Space-Air-Ground Integrated Network: A Survey},'' \emph{{IEEE} Commun. Surveys Tuts}, vol.~20, no.~4, pp. 2714--2741, 2018.

\bibitem{cui2022space}
H.~Cui, J.~Zhang, Y.~Geng, Z.~Xiao, T.~Sun, N.~Zhang, J.~Liu, Q.~Wu, and X.~Cao, ``{Space-Air-Ground Integrated Network (SAGIN) for 6G: Requirements, Architecture And Challenges},'' \emph{China Commun}, vol.~19, no.~2, pp. 90--108, 2022.

\bibitem{yue2023low}
P.~Yue, J.~An, J.~Zhang, J.~Ye, G.~Pan, S.~Wang, P.~Xiao, and L.~Hanzo, ``{Low Earth Orbit Satellite Security and Reliability: Issues, Solutions, and the Road Ahead},'' \emph{{IEEE} Commun. Surveys Tuts}, 2023.

\bibitem{lin2021spatial}
Z.~Lin, L.~Wang, B.~Tan, and X.~Li, ``Spatial-spectral terahertz networks,'' \emph{IEEE Transactions on Wireless Communications}, vol.~21, no.~6, pp. 3881--3892, 2021.

\bibitem{whitepaper}
M.~G. T.~W. Paper, \emph{{6G Satellite and Terrestrial Network Convergence}}.\hskip 1em plus 0.5em minus 0.4em\relax Cambridge university press, 2023.

\bibitem{chakraborty2017specsense}
A.~Chakraborty, M.~S. Rahman, H.~Gupta, and S.~R. Das, ``{Specsense: Crowdsensing for Efficient Querying of Spectrum Occupancy},'' in \emph{Proc. INFOCOM}, 2017, pp. 1--9.

\bibitem{uvaydov2021deepsense}
D.~Uvaydov, S.~D’Oro, F.~Restuccia, and T.~Melodia, ``{Deepsense: Fast Wideband Spectrum Sensing Through Real-Time in-the-Loop Deep Learning},'' in \emph{Proc. INFOCOM}, 2021, pp. 1--10.

\bibitem{luo2023spectrum}
Z.~Luo, Q.~Huang, X.~Chen, R.~Wang, F.~Wu, G.~Chen, and Q.~Zhang, ``{Spectrum Sensing Everywhere: Wide-Band Spectrum Sensing With Low-Cost UWB Nodes},'' \emph{{IEEE/ACM} Trans. Netw}, 2023.

\bibitem{peng2024sums}
J.~Peng, Z.~Chen, Z.~Lin, H.~Yuan, Z.~Fang, L.~Bao, Z.~Song, Y.~Li, J.~Ren, and Y.~Gao, ``Sums: Sniffing unknown multiband signals under low sampling rates,'' \emph{IEEE Transactions on Mobile Computing}, 2024.

\bibitem{bi2019engineering}
S.~Bi, J.~Lyu, Z.~Ding, and R.~Zhang, ``{Engineering Radio Maps for Wireless Resource Management},'' \emph{{IEEE} Wireless Commun}, vol.~26, no.~2, pp. 133--141, 2019.

\bibitem{zeng2024tutorial}
Y.~Zeng, J.~Chen, J.~Xu, D.~Wu, X.~Xu, S.~Jin, X.~Gao, D.~Gesbert, S.~Cui, and R.~Zhang, ``{A Tutorial on Environment-Aware Communications via Channel Knowledge Map for 6G},'' \emph{IEEE Communications Surveys \& Tutorials}, 2024.

\bibitem{shen20213d}
F.~Shen, Z.~Wang, G.~Ding, K.~Li, and Q.~Wu, ``{3D Compressed Spectrum Mapping with Sampling Locations Optimization in Spectrum-Heterogeneous Environment},'' \emph{{IEEE} Trans. Wireless Commun}, vol.~21, no.~1, pp. 326--338, 2021.

\bibitem{wang2023sparse}
J.~Wang, Q.~Zhu, Z.~Lin, Q.~Wu, Y.~Huang, X.~Cai, W.~Zhong, and Y.~Zhao, ``{Sparse Bayesian Learning-Based 3D Radio Environment Map Construction—Sampling Optimization, Scenario-Dependent Dictionary Construction and Sparse Recovery},'' \emph{{IEEE} Trans. on Cogn. Commun. Netw}, 2023.

\bibitem{liu2024uav}
C.~Liu, K.~Zhu, C.~Tao, B.~Chen, and Y.~Zhao, ``{UAV-Assisted Active Sparse Crowdsensing for Ground Signal Map Construction Based on 3-D Spatial-Temporal Correlation},'' \emph{{IEEE} Internet Things J}, 2024.

\bibitem{hu20233d}
T.~Hu, Y.~Huang, J.~Chen, Q.~Wu, and Z.~Gong, ``{3D Radio Map Reconstruction Based on Generative Adversarial Networks Under Constrained Aircraft Trajectories},'' \emph{{IEEE} Trans. Veh}, vol.~72, no.~6, pp. 8250--8255, 2023.

\bibitem{zhang2020spectrum}
G.~Zhang, X.~Fu, J.~Wang, X.-L. Zhao, and M.~Hong, ``{Spectrum cartography via coupled block-term tensor decomposition},'' \emph{{IEEE} Trans. Signal Process}, vol.~68, pp. 3660--3675, 2020.

\bibitem{shrestha2022deep}
S.~Shrestha, X.~Fu, and M.~Hong, ``{Deep Spectrum Cartography: Completing Radio Map Tensors Using Learned Neural Models},'' \emph{{IEEE} Trans. Signal Process}, vol.~70, pp. 1170--1184, 2022.

\bibitem{teganya2021deep}
Y.~Teganya and D.~Romero, ``{Deep Completion Autoencoders for Radio Map Estimation},'' \emph{{IEEE} Trans. Wireless Commun}, vol.~21, no.~3, pp. 1710--1724, 2021.

\bibitem{guddeti2019sweepsense}
Y.~Guddeti, R.~Subbaraman, M.~Khazraee, A.~Schulman, and D.~Bharadia, ``{$\{$SweepSense$\}$: Sensing 5 $\{$GHz$\}$ in 5 Milliseconds with Low-cost Radios},'' in \emph{Proc. NSDI}, 2019, pp. 317--330.

\bibitem{humphreys2023signal}
T.~E. Humphreys, P.~A. Iannucci, Z.~M. Komodromos, and A.~M. Graff, ``{Signal Structure of the Starlink Ku-Band Downlink},'' \emph{{IEEE} Trans. Aerosp. Electron. Syst}, vol.~59, no.~5, pp. 6016--6030, 2023.

\bibitem{mcdowell2020low}
J.~C. McDowell, ``{The Low Earth Orbit Satellite Population and Impacts of the SpaceX Starlink Constellation},'' \emph{The Astrophysical Journal Letters}, vol. 892, no.~2, p. L36, 2020.

\bibitem{song2019real}
Z.~Song, H.~Qi, and Y.~Gao, ``{Real-Time Multi-Gigahertz Sub-Nyquist Spectrum Sensing System for mmwave},'' in \emph{Proceedings of the 3rd ACM Workshop on Millimeter-wave Networks and Sensing Systems}, 2019, pp. 33--38.

\bibitem{yang2019fast}
L.~Yang, J.~Fang, H.~Duan, and H.~Li, ``{Fast compressed power spectrum estimation: Toward a practical solution for wideband spectrum sensing},'' \emph{{IEEE} Trans. Wireless Commun}, vol.~19, no.~1, pp. 520--532, 2019.

\bibitem{xu2013parallel}
Y.~Xu, R.~Hao, W.~Yin, and Z.~Su, ``{Parallel Matrix Factorization for Low-Rank Tensor Completion},'' \emph{arXiv preprint arXiv:1312.1254}, 2013.

\bibitem{xie2017kronecker}
Q.~Xie, Q.~Zhao, D.~Meng, and Z.~Xu, ``{Kronecker-Basis-Representation Based Tensor Sparsity and Its Applications to Tensor Recovery},'' \emph{{IEEE} Trans. Pattern Anal. Mach. Intell}, vol.~40, no.~8, pp. 1888--1902, 2017.

\bibitem{perenda2021learning}
E.~Perenda, S.~Rajendran, G.~Bovet, S.~Pollin, and M.~Zheleva, ``{Learning the Unknown: Improving Modulation Classification Performance in Unseen Scenarios},'' in \emph{Proc. INFOCOM}, 2021, pp. 1--10.

\bibitem{perenda2023contrastive}
E.~Perenda, S.~Rajendran, G.~Bovet, M.~Zheleva, and S.~Pollin, ``{Contrastive learning with self-reconstruction for channel-resilient modulation classification},'' in \emph{Proc. INFOCOM}, 2023, pp. 1--10.

\bibitem{scalingi2024det}
A.~Scalingi, S.~D'Oro, F.~Restuccia, T.~Melodia, D.~Giustiniano \emph{et~al.}, ``{Det-RAN: Data-Driven Cross-Layer Real-Time Attack Detection in 5G Open RANs},'' in \emph{Proc. INFOCOM}, 2024, pp. 1--10.

\bibitem{lin2024split}
Z.~Lin, G.~Qu, X.~Chen, and K.~Huang, ``Split learning in 6g edge networks,'' \emph{IEEE Wireless Communications}, 2024.

\bibitem{feng2023dynamic}
M.~Feng, W.~Zhang, and M.~Krunz, ``{Dynamic Spectrum Access in Non-stationary Environments: A DRL-LSTM Integrated Approach},'' in \emph{Proc. ICNC}, 2023, pp. 159--164.

\bibitem{sanchez2022airnn}
S.~G. Sanchez and S.~I. et~al., ``{AirNN: Over-the-Air Computation for Neural Networks via Reconfigurable Intelligent Surfaces},'' \emph{{IEEE/ACM} Trans. Netw}, vol.~31, no.~6, pp. 2470--2482, 2022.

\bibitem{zhang2021signal}
W.~Zhang, M.~Feng, M.~Krunz, and A.~H.~Y. Abyaneh, ``{Signal Detection and Classification in Shared Spectrum: A Deep Learning Approach},'' in \emph{Proc. INFOCOM}, 2021, pp. 1--10.

\bibitem{lin2024efficient}
Z.~Lin, G.~Zhu, Y.~Deng, X.~Chen, Y.~Gao, K.~Huang, and Y.~Fang, ``Efficient parallel split learning over resource-constrained wireless edge networks,'' \emph{IEEE Transactions on Mobile Computing}, 2024.

\bibitem{qayyum2022untrained}
A.~Qayyum, I.~Ilahi, F.~Shamshad, F.~Boussaid, M.~Bennamoun, and J.~Qadir, ``{Untrained Neural Network Priors for Inverse Imaging Problems: A Survey},'' \emph{{IEEE} Trans. Pattern Anal. Mach. Intell}, vol.~45, no.~5, pp. 6511--6536, 2022.

\bibitem{luo2023low}
Y.~Luo, X.~Zhao, Z.~Li, M.~K. Ng, and D.~Meng, ``{Low-Rank Tensor Function Representation for Multi-Dimensional Data Recovery},'' \emph{{IEEE} Trans. Pattern Anal. Mach. Intell}, 2023.

\bibitem{fan2021multi}
J.~Fan, ``{Multi-Mode Deep Matrix and Tensor Factorization},'' in \emph{Proc. ICLR}, 2021.

\bibitem{yang2021adaptive}
J.~Yang, Z.~Song, Y.~Gao, X.~Gu, and Z.~Feng, ``{Adaptive Compressed Spectrum Sensing for Multiband Signals},'' \emph{{IEEE} Trans. Wireless Commun}, vol.~20, no.~11, pp. 7642--7654, 2021.

\bibitem{li2023lightnestle}
Y.~Li, W.~Liang, K.~Xie, D.~Zhang, S.~Xie, and K.~Li, ``{Lightnestle: Quick and Accurate Neural Sequential Tensor Completion via Meta Learning},'' in \emph{Proc. INFOCOM}, 2023, pp. 1--10.

\bibitem{3GPP2020}
\BIBentryALTinterwordspacing
3GPP. (2020) {Study on New Radio (NR) to Support Non-Terrestrial Networks}. 3rd Generation Partnership Project (3GPP). [Online]. Available: \url{https://www.3gpp.org/ftp/Specs/archive/38_series/38.811/}
\BIBentrySTDinterwordspacing

\bibitem{goldsmith2005wireless}
A.~Goldsmith, \emph{{Wireless Communications}}.\hskip 1em plus 0.5em minus 0.4em\relax Cambridge university press, 2005.

\bibitem{fang2016two}
J.~Fang, L.~Zhang, and H.~Li, ``{Two-Dimensional Pattern-Coupled Sparse Bayesian Learning via Generalized Approximate Message Passing},'' \emph{{IEEE} Trans. Image Process}, vol.~25, no.~6, pp. 2920--2930, 2016.

\bibitem{setianto2013comparison}
A.~Setianto and T.~Triandini, ``{Comparison of Kriging and Inverse Distance Weighted (IDW) Interpolation Methods in Lineament Extraction and Analysis},'' \emph{Journal of Applied Geology}, vol.~5, no.~1, 2013.

\bibitem{chen2020robust}
L.~Chen, X.~Jiang, X.~Liu, and Z.~Zhou, ``{Robust Low-Rank Tensor Recovery via Nonconvex Singular Value Minimization},'' \emph{{IEEE} Trans. Image Process}, vol.~29, pp. 9044--9059, 2020.

\bibitem{mishali2009blind}
M.~Mishali and Y.~C. Eldar, ``{Blind Multiband Signal Reconstruction: Compressed Sensing for Analog Signals},'' \emph{{IEEE} Trans. Signal Process}, vol.~57, no.~3, pp. 993--1009, 2009.

\bibitem{song2022approaching}
Z.~Song, J.~Yang, H.~Zhang, and Y.~Gao, ``{Approaching Sub-Nyquist Boundary: Optimized Compressed Spectrum Sensing Based on Multicoset Sampler for Multiband Signal},'' \emph{{IEEE} Trans. Signal Process}, vol.~70, pp. 4225--4238, 2022.

\bibitem{chen2022constrained}
X.~Chen, J.~Wang, Q.~Peng, and G.~Zhang, ``{A Constrained Block-Term Tensor Decomposition Framework for Spectrum Cartography},'' \emph{{IEEE} Signal Process. Lett}, vol.~29, pp. 1699--1703, 2022.

\bibitem{zhang2024fedac}
Y.~Zhang, H.~Chen, Z.~Lin, Z.~Chen, and J.~Zhao, ``Fedac: A adaptive clustered federated learning framework for heterogeneous data,'' \emph{arXiv preprint arXiv:2403.16460}, 2024.

\bibitem{lin2024adaptsfl}
Z.~Lin, G.~Qu, W.~Wei, X.~Chen, and K.~K. Leung, ``Adaptsfl: Adaptive split federated learning in resource-constrained edge networks,'' \emph{arXiv preprint arXiv:2403.13101}, 2024.

\bibitem{hu2024accelerating}
M.~Hu, J.~Zhang, X.~Wang, S.~Liu, and Z.~Lin, ``Accelerating federated learning with model segmentation for edge networks,'' \emph{IEEE Transactions on Green Communications and Networking}, 2024.

\bibitem{leo2025split}
Z.~Lin, Y.~Zhang, Z.~Chen, Z.~Fang, C.~Wu, X.~Chen, Y.~Gao, and J.~Luo, ``{LEO-Split: A Semi-Supervised Split Learning Framework over LEO Satellite Networks},'' \emph{arXiv preprint arXiv:2501.01293}, Jan. 2025.

\bibitem{zhang2024satfed}
Y.~Zhang, Z.~Lin, Z.~Chen, Z.~Fang, W.~Zhu, X.~Chen, J.~Zhao, and Y.~Gao, ``Satfed: A resource-efficient leo satellite-assisted heterogeneous federated learning framework,'' \emph{arXiv preprint arXiv:2409.13503}, 2024.

\bibitem{lin2024hierarchical}
Z.~Lin, W.~Wei, Z.~Chen, C.-T. Lam, X.~Chen, Y.~Gao, and J.~Luo, ``Hierarchical split federated learning: Convergence analysis and system optimization,'' \emph{arXiv preprint arXiv:2412.07197}, 2024.

\bibitem{zhang2025lcfed}
Y.~Zhang, H.~Chen, Z.~Lin, Z.~Chen, and J.~Zhao, ``{LCFed: An Efficient Clustered Federated Learning Framework for Heterogeneous Data},'' \emph{arXiv preprint arXiv:2501.01850}, Jan. 2025.

\bibitem{lin2023pushing}
Z.~Lin, G.~Qu, Q.~Chen, X.~Chen, Z.~Chen, and K.~Huang, ``Pushing large language models to the 6g edge: Vision, challenges, and opportunities,'' \emph{arXiv preprint arXiv:2309.16739}, 2023.

\bibitem{hu2024agentscodriver}
S.~Hu, Z.~Fang, Z.~Fang, Y.~Deng, X.~Chen, and Y.~Fang, ``{AgentsCoDriver: Large Language Model Empowered Collaborative Driving with Lifelong Learning},'' \emph{arXiv preprint arXiv:2404.06345}, Apr. 2024.

\bibitem{fang2024automated}
Z.~Fang, Z.~Lin, Z.~Chen, X.~Chen, Y.~Gao, and Y.~Fang, ``Automated federated pipeline for parameter-efficient fine-tuning of large language models,'' \emph{arXiv preprint arXiv:2404.06448}, 2024.

\bibitem{hu2021lora}
E.~J. Hu, Y.~Shen, P.~Wallis, Z.~Allen-Zhu, Y.~Li, S.~Wang, L.~Wang, and W.~Chen, ``Lora: Low-rank adaptation of large language models,'' \emph{arXiv preprint arXiv:2106.09685}, 2021.

\bibitem{hu2024agentscomerge}
S.~Hu, Z.~Fang, Z.~Fang, Y.~Deng, X.~Chen, Y.~Fang, and S.~Kwong, ``{AgentsCoMerge: Large Language Model Empowered Collaborative Decision Making for Ramp Merging},'' \emph{arXiv preprint arXiv:2408.03624}, Aug. 2024.

\bibitem{lin2024splitlora}
Z.~Lin, X.~Hu, Y.~Zhang, Z.~Chen, Z.~Fang, X.~Chen, A.~Li, P.~Vepakomma, and Y.~Gao, ``Splitlora: A split parameter-efficient fine-tuning framework for large language models,'' \emph{arXiv preprint arXiv:2407.00952}, 2024.

\end{thebibliography}

\vfill

\end{document}